\documentclass{aa}

\usepackage[varg]{txfonts}
\usepackage{natbib}
\usepackage{graphicx}
\usepackage{times}
\usepackage{color}
\usepackage[colorlinks=true, citecolor=blue, linkcolor=black, urlcolor=black]{hyperref}
\usepackage{txfonts}
\usepackage{placeins}           

\bibpunct{(}{)}{;}{a}{}{,}
\newcommand{\lat}{$\lambda$~Tau }
\newcommand{\lae}{$\lambda$~Tau}

\newcommand{\respefoe}{{\tt reSPEFO}}
\newcommand{\phoebe}{{\tt PHOEBE} }

\newcommand{\fotel}{{\tt FOTEL} }
\newcommand{\fotele}{{\tt FOTEL}}
\newcommand{\korel}{{\tt KOREL} }
\newcommand{\korele}{{\tt KOREL}}

\newcommand{\ubv}{\hbox{$U\!B{}V$}}
\newcommand{\bv}{\hbox{$B\!-\!V$}}
\newcommand{\bvr}{\hbox{$B{}VR$}}
\newcommand{\ub}{\hbox{$U\!-\!B$}}

\newcommand{\ubvr}{\hbox{$U\!B{}V\!R$}}
\newcommand{\hp}{\hbox{$H_{\rm p}$}}
\newcommand{\oc}{\hbox{$O\!-\!C$}}

\newcommand{\m}{$^{\rm m}\!\!.$}

\newcommand{\kms}{km~s$^{-1}$ }
\newcommand{\ks}{km~s$^{-1}$}
\newcommand{\vsin}{$v$~sin~$i$ }

\newcommand{\ms}{M$_{\odot}$}
\newcommand{\rs}{R$_{\odot}$}

\newcommand{\ha}{H$\alpha$ }

\newcommand{\hae}{H$\alpha$}

\newcommand{\Ame}{\AA~mm$^{-1}$}

\newcommand{\n}{\hbox{\v{n}}}

\begin{document}

   \title{Towards a new model of the compact triple system $\lambda$~Tauri\\
   I. Improved periods and preliminary elements of the triple system}

\author{P.~Harmanec\inst{\ref{praha}}\corrauth{Petr.Harmanec@matfyz.cuni.cz}\and
J.~\v{S}vr\v{c}kov\'a\inst{\ref{praha}}\and
P.~Dole\v{z}al\inst{\ref{praha}}\and
H.~Bo\v{z}i\'c\inst{\ref{hvar}}\and
D.~Vokrouhlick\'y\inst{\ref{praha}}\and
P.~Zasche\inst{\ref{praha}}\and
M.~\v{S}lechta\inst{\ref{ondr}}\and
B.N.~Barlow\inst{\ref{chapelhill}}\and
T.~Pribulla\inst{\ref{tatry}}\and
T.~Kallinger\inst{\ref{vienna}}    \and 
R.~Kuschnig\inst{\ref{graz}}       \and 
J.~Matthews\inst{\ref{vancouver}}  \and 
W.W.~Weiss\inst{\ref{vienna}}      \and 
D.~Kor\v{c}\'akov\'a\inst{\ref{praha}}\and
M.~Wolf\inst{\ref{praha}}\and
D.~Mourard\inst{\ref{nice}}\and
N.~Nardetto\inst{\ref{nice}}\and
D.~Terrell\inst{\ref{boulder}}\and
G.~Pietrzynski\inst{\ref{warsaw}}\and
J.~Ribeiro\inst{\ref{lisboa}}\and
R.~Uhl\'a\v{r}\inst{\ref{jilove}}\and
R.~K\v{r}\'\i\v{c}ek\inst{\ref{praha}}
}

   \institute{
   Astronomical Institute of Charles University,
   Faculty of Mathematics and Physics,\hfill\break
   V~Hole\v{s}ovi\v{c}k\'ach~2, CZ-180 00 Praha~8 - Troja,
   Czech Republic\label{praha}
\and
    Hvar Observatory, Faculty of Geodesy, University of Zagreb,
    HR-10000 Zagreb, Croatia\label{hvar}
\and
    Astronomical Institute, Czech Academy of Sciences, CZ-25165 Ond\v{r}ejov,
    Czech Republic\label{ondr}
\and
    Astronomical Institute of the Slovak Academy of Sciences,
    059 60 Tatransk\'a Lomnica, The Slovak Republic\label{tatry}
\and
    Laboratoire Lagrange, UMR7293, Universit\'e de Nice
    Sophia-Antipolis, CNRS, Observatoire de la C\^ote d'Azur,
    Nice, France\label{nice}
\and
    Universit\"at Graz, Institut f{\"u}r Physik, Universit\"atsplatz 3,
    8010 Graz, Austria\label{graz}
\and
    Universit\"at Wien, Institut f{\"u}r Astrophysik, T{\"u}rkenschanzstrasse 17,
    1180 Vienna, Austria\label{vienna}
\and
    University of British Columbia, Department of Physics and Astronomy,
    6224 Agricultural Road, Vancouver, BC V6T 1Z1, Canada\label{vancouver}
\and
    Nicolaus Copernicus Astronomical Center, Warsaw, Poland\label{warsaw}
\and
    Department of Space Studies, Southwest Research Institute,
    Boulder, CO 80302, USA\label{boulder}
\and
 Department of Physics and Astronomy, University of North Carolina at
 Chapel Hill, Chapel Hill, NC 27599, USA \label{chapelhill}
\and
    Observat\'{o}rio Astron\'{o}mico do Centro de Informa\c{c}\~{a}o Geoespacial
    do Ex\'{e}rcito, Lisboa, Portugal\label{lisboa}
 \and
    Private Observatory, Poho\v{r}\'\i~71, CZ-254~01~J\'\i lov\'e u Prahy,
    Czech Republic\label{jilove}
  }

\date{Received \today}

\abstract{We analysed our new radial velocity
 measurements and photometric observations (covering a time interval of
 46838~days and 39100~days, respectively) as well as those from previous studies to derive accurate ephemerides
 for both orbits of the compact triple system \lae. In the case of
 the inner eclipsing subsystem, we obtained a~linear ephemeris of
$T_{\rm min.I}={\rm HJD}\,2440998.86069(35)+3\fd952952845(81)\cdot E$, improving the accuracy of the previous solutions by more than an order of
magnitude. This allowed us to reproduce well all available light curve minima,
including the first one recorded in 1844. Furthermore, the period of the outer subsystem,
determined from radial velocity analysis, appears to be constant over the interval
of available observations, with the corresponding linear ephemeris being
$T_{\rm periast.}={\rm HJD}\,2456822.5(3.6)+33\fd02480(28)\cdot E\,$.
Assuming near coplanarity of the inner and outer orbits, we estimated the
preliminary physical properties of the whole system. We found that the
secondary of the inner subsystem is filling or nearly filling its Roche
lobe. We used the long-term stability of the eclipsing subsystem period to place a
strong upper limit on the rate of the proposed episodic mass transfer to its
primary component. Our results constitute a preparatory step for a follow-up
study that will also include an analysis of a rich collection of optical
spectro-interferometric observations and use a full-fledged numerical model
to describe the evolution of the orbital architecture of the system.
Particular attention will be focused on determining the inner orbit
forced eccentricity and understanding the absence of the apsidal
advance of the outer orbit in our present solution.}

\keywords{Stars: close --
          Stars: binaries: spectroscopic --
          Stars: binaries: eclipsing --
          Stars: fundamental parameters --
          Stars: individual: \lae}

   \titlerunning{Improved properties of \lae}
\maketitle
\nolinenumbers

\section{Introduction}\label{one}
Compact triple systems are extremely interesting objects in stellar astrophysics, as
they represent unique laboratories to test various hypotheses
related to their origin, they are subject to a plethora of complicated
evolutionary pathways, and they offer a possibility of revealing traces of
dynamical effects of various timescales within affordable decade-long
periods. The latter, if properly modelled, consequently provides a unique tool
to constrain the physical parameters of the stars. Analysis of the
data from \textit{Kepler} and the Transiting Exoplanet Survey Satellite (TESS)
space-borne surveys has provided spectacular
examples of systems in this category. For example, \citet{kostov2024} reported
the discovery of the so-far most compact triply eclipsing triple system,
TIC~290061484 = 2MASS~J21042626+5113320, with an outer period of
24\fd5 only, and \citet{borko2025} reported ten new compact systems with outer
periods between 46\fd8 and 101\fd4.
Even more recently, \citet{rappaport2026} discovered TIC~366240660,
a~triply eclipsing system with an outer orbit period of 30\fd5 in a rigorous
13:1 mean motion resonant configuration with the inner orbit.
Although these new systems are fascinating in many
respects and have brought evidence of previously unseen configurations,
\lat -- with an outer orbit period of only 33~d -- remains the prime and
archetypal representative of a stellar compact triple hierarchy. Among other
reasons, its uniqueness
stems from its brightness and its long observational history. Considering
the availability of numerous spectroscopic and photometric observations
of \lae ~(briefly overviewed in Sect.~2) and reporting results from
our own dedicated campaign, which provide a new dense series of spectroscopic and
photometric observations, the goal of the present paper is to provide
reliable and accurate ephemerides for both inner and outer orbits in the
\lae ~system. These will serve as a starting point
for a follow-up study in which we will make use of additional
interferometric observations and construct a state-of-the-art dynamical model
of this remarkable triple star.

\section{Relevant observational facts about \lae}\label{two}
The star \lat (HD~25204, HR~1239, HIP~18724) is one of the brightest
eclipsing binaries in the northern sky
\citep[$V$=3\m4, \bv=$-0$\m12, \ub=$-$0.63;][]{john66}.
Its light variations were reported as early as 1848 by \citet{bax1848},
who already realised that the period of light changes must be close to 4~days.
Later, it turned out that the very first minimum was in fact already observed by
\citet{schmidt1858} on January 21, 1844.
The first photoelectric light curve was published by \citet{steb20}, who
observed the star relative to
$\xi$~Tau, $\pi^4$~Ori, $\pi^5$~Ori, $\eta$~Tau, and $\mu$~Tau. He discovered
\citep[see also][]{steb20b} that $\pi^5$~Ori is an ellipsoidal
variable with a period of 3\fd70045 and a full amplitude of 0\m06. Using
a~green-sensitive diode, Stebbins observed \lat through an attenuating
filter, carefully calibrated its transparency, compensated also for the
variations of $\pi^5$~Ori, and finally reduced all observations relative
to $\xi$~Tau. Interestingly enough, $\xi$~Tau is now also known to be an
eclipsing binary with a 7\fd15 period
\citep{bolt2007,nemr2013,nemr2016,broz2026}. Stebbins's
observations defined the eclipsing light curve of \lat quite well and allowed
him to derive the ephemeris for the primary minimum:
\begin{equation}
T_{\rm min.I}={\rm HJD}\,2421506.850+3\fd952952\cdot E\,.\label{efe1}
\end{equation}
\noindent Another (blue) light curve of \lat was obtained by \citet{nik38}
relative to $\mu$~Tau and $\nu$~Tau.
The first \ubv\ light curve of \lat was obtained by \citet{grant59} relative
to HR~1194. His comparison star also turned out to be a light variable
(of CP2 type) with a period of 2\fd5345 and a range of 0\m06 in Str\"omgren $y$
\citep{rens79,wra2012} and is now known as V766~Tau.

The changes in radial velocity (RV) of \lat were first reported by
\citet{belo1898}, and the first RV curve with a period of 3\fd952941,
based on 23 Potsdam spectra, was published by \citet{aret13}. The presence
of a close tertiary moving around the centre of mass of the eclipsing binary
with a~period of 34\fd6 was discovered by \citet{schle16a, schle16b}.
\citet{mcl37} obtained a~new series of spectra and concluded that a better value
for the long period is 30\fd0. \citet{ebbig56} obtained an~additional series of
spectrograms at several observatories and analysed them together with
previously published ones to arrive at a period of 33\fd025 for the long
orbit. They noted that the \citet{steb20} ephemeris (\ref{efe1}) for the short
orbit satisfies all available RVs. Later, \citet{grant59} studied 20 coud\'e
spectra
from the McDonald 82-inch reflector and demonstrated the presence of the
weak lines of the secondary in the \ha and \ion{Mg}{II}~4481~\AA\ lines.
He did not publish any RVs but estimated the component masses to be
6.06 and 1.61~\ms. \citet{casi68} measured RVs on 16 photographic grating
spectra with a dispersion of 34~\Ame\ and derived new orbital elements.
However, they did not publish the individual RVs. \citet{fekel82} obtained
the first electronic (Reticon) spectra of \lat covering the wavelength region
from 4460 to 4515~\AA, and they
were able to detect the signal of the secondary in several metallic lines and
measure their RVs. They found that the RV residuals from the 3\fd95 orbit
follow the 33\fd ~orbit for both primary and secondary RVs. They also provided
some arguments in favour of near coplanarity of both orbits.
\citet{berdy2018} obtained 69 \bvr\ polarimetric observations of \lat between
December 2012 and November 2016 at two observatories.
They found small periodic changes with double-wave phase curves in the normalised
Stokes parameters $q$ and $u$ in all three photometric passbands.
Using an~analytical method, they estimated the orbital inclination of the inner eclipsing
binary to be between about 74$^\circ$ and $77^\circ$, while for scattering
models, they found values between $72^\circ$ and $76^\circ$. They confirmed
that the inner 3\fd95 system is a~semi-detached one, with the cooler secondary
filling its Roche lobe. They also found that the revolution of the inner system,
as seen on the plane of the sky, is clockwise.

\section{Observational material used}\label{three}

Throughout this paper, we express the times of observations
in the reduced Julian date defined as

\smallskip
\centerline{RJD = HJD$-2400000.0$\ ,}

\noindent with the exception of very old photometry obtained before JD~2400000.
 It is true that it would be better to use the barycentric
Julian dates (BJDs) consistently with all effects of non-uniform Earth rotation included.
However, for many of the datasets we used, the accurate times
of mid-exposures are not known, and only Julian dates to three decimal digits
are provided. Since the differences between HJDs and BJDs in days are only in
the fourth or fifth decimal place, it is impossible to calculate BJDs precisely 
enough for such data. Doing so is only possible for more recent datasets, and
we postpone this analysis to the follow-up study.

\subsection{Spectroscopy}

\begin{table}
\caption[]{Published available RV observations of $\lambda$ Tau.}
\label{jourv}
\begin{flushleft}
\begin{tabular}{rccll}
\hline\hline\noalign{\smallskip}
Spg.&Time interval&   No. of RVs    &Source\\
No. &         (RJD) &prim./sec.& \\
\noalign{\smallskip}\hline\noalign{\smallskip}
 1&14230.47--14268.38&  5/2&A\\
 2&15689.50--15849.27& 23/0&B\\
 3&17830.82--20210.51& 28/0&C\\
 4&17523.77--20915.53&107/0&D\\
 5&23743.70--24157.67& 91/0&E\\
 6&28048.01--29295.62& 36/0&D\\
 7&34326.75--35143.78& 68/0&D\\
 8&35017.01--35036.80&  2/0&D\\
 9&44922.72          &  1/0&F\\
10&44177.90--44955.89&44/44&G\\
\noalign{\smallskip}\hline
\end{tabular}
\tablefoot{Spectrographs and telescopes in column ``Spg. No.":\\
1... Pulkowa 0.75~m refractor, 2-prism spg. 14~\Ame;\\
2... Potsdam 0.80~m refractor, prism spg., 27~\Ame\ at H$\gamma$, modern
     wavelengths;\\
3... Allegheny 0.79~m reflector, 1-prism spg., 40~\Ame, old wavelengths;\\
4... Allegheny 0.79~m reflector, 1-prism spg., 40~\Ame, modern wavelengths;\\
5... Ann Arbor 0.95~m reflector, 1-prism spg., 40~\Ame\ at H$\gamma$;\\
6... Dominion Astrophysical Observatory 1.83~m reflector, prism spg.;\\
7... MtWilson 1.52~m reflector, prism. spg., 36~\Ame  at H$\gamma$;\\
8...Lick 0.91~m refractor, Mills 3-prism spg.;\\
9...Kitt Peak Observatory coud\'e auxiliary 0.90~m reflector, grating spg.
     16.9~\Ame;\\
10... McDonald 2.7~m reflector, grating. Reticon spg.;\\
Abbreviations in column ``Source" stand for:\\
A... \citet{belo1898};
B... \citet{aret13};
C... \citet{schle16a,schle16b};
D... \citet{ebbig56};
E... \citet{mcl37};
F... \citet{zen83};
G... \citet{fekel82};
}
\end{flushleft}
\end{table}

\begin{table}
\caption[]{Available electronic spectra of $\lambda$ Tau.}
\label{jousp}
\begin{flushleft}
\begin{tabular}{rcrcrcll}
\hline\hline\noalign{\smallskip}
Spg.&Time interval&No. of&Wavelength&Spectral \\
No. &       (RJD) &sp.&range (\AA)&res. \\
\noalign{\smallskip}\hline\noalign{\smallskip}
11&54741.80--55501.89& 25&6320--6710& 60000\\
12&54807.56--56564.75& 12&3700--9220& 48000\\
13&55060.87--56578.90& 17&6260--6910&115000\\
14&56154.62--58390.53& 64&6260--6740& 11700\\
14&56235.36--57075.25&  5&4270--4510& 11700\\
15&57354.38--58075.39& 17&6520--6700& 14000\\
16&59472.87--59504.85&165&4500--7600& 79000\\
16&59865.70--59911.76&250&4500--7600& 79000\\
17&60622.57--60729.34& 13&4240--7260&$\sim$30000\\
17&60920.58--61068.40& 20&4240--7260&$\sim$30000\\
\noalign{\smallskip}\hline
\end{tabular}
\tablefoot{Spectrographs and telescopes in column ``Spg. No.":\\
11... ESO LaSilla Leonard Euler Nasmyth 1.2~m reflector,
      Coralie echelle CCD spectrograph;\\
12... ESO LaSilla 2.2 m Cassegrain reflector,
      FEROS echelle CCD spectrograph;\\
13... ESO LaSilla 3.6 m reflector, HARPS echelle CCD spectrograph;\\
14... Ond\v{r}ejov 2.0 m reflector, coud\'e grating SITe-5 CCD spg.;\\
15... Lisboa IGeoE 0.356 m SC reflector, Littrow LHIRES\,III spg.;\\
16... CTIO 1.5 m reflector, CHIRON echelle spectrograph;\\
17... Skalnat\'e Pleso 1.3 m Nasmyth-Cassegrain reflector,
      echelle spectrograph.
}
\end{flushleft}
\end{table}

We compiled and carefully re-examined all published RV measurements of
\lae. Their log is provided in Table~\ref{jourv} and details on
the individual datasets are in Appendix~\ref{apa}.
Our new observational material consists of the following series
of electronic spectra from five observatories secured with seven
different spectrographs: Red segments of 25 usable ESO La Silla 
Coralie CCD spectra; ESO La Silla FEROS echelle CCD spectra, one 
from 2008 and eleven from a~close series in 2013;
ESO La Silla HARPS echelle CCD spectra;
Ond\v{r}ejov 64 red and five blue CCD spectra;
Lisboa 0.356~m Littrow CCD spectra; CHIRON echelle spectra from 
Cerro Tololo Inter-American Observatory (CTIO); and echelle spectra 
from the Skalnat\'e Pleso Observatory.
A complete log of these electronic spectra is given in Table~\ref{jousp}
and the details on the spectra reduction and RV measurements can also be
found in Appendix~\ref{apa}.

\subsection{Photometry}

\begin{table*}
\caption[]{Available photometric observations of \lae.}\label{jouphot}
\begin{flushleft}
\begin{tabular}{ccrccll}
\hline\noalign{\smallskip}
File&Time interval& No. of &Passbands&HD of comparison&Source\\
       &(HJD)&obs.  &  &/ check star\\
\noalign{\smallskip}\hline
\hline\noalign{\smallskip}
A&2396338.4--2396354.4&              5& visual& 25490/many&\citet{bax1848}  \\
B&2421159.8--2421635.5&        150/145&pe-cell&      21364&\citet{steb20}   \\
B&2421164.9--2421321.6&             24&pe-cell&      31237&\citet{steb20}   \\
B&2421202.7--2421296.5&          16/15&pe-cell&      30836&\citet{steb20}   \\
B&2421482.9--2421635.6&          56/55&pe-cell&      23793&\citet{steb20}   \\
B&2421514.9--2421670.6&             21&pe-cell&      26912&\citet{steb20}   \\
C&2428057.5--2428115.6&             48&blue pe&26912/25490&\citet{nik38}    \\
C&2428421.5--2428484.4&             46&blue pe&26912/25490&\citet{nik38}    \\
C&2428788.5--2428816.6&             38&blue pe&26912/25490&\citet{nik38}    \\
D&2435091.6--2435160.7&    228/230/310& \ubv  &24155/26171&\citet{grant59}  \\
E&2447911.7--2448828.7&             94& \hp   &        -- &\citet{esa97}\\
F&2454216.1--2454232.3&            481& STEREO&        -- &this paper\\
F&2454251.5--2454273.9&            497& STEREO&        -- &this paper\\
F&2454559.9--2454576.4&            525& STEREO&        -- &this paper\\
F&2454642.7--2454661.8&            464& STEREO&        -- &this paper\\
F&2454903.7--2454922.3&            561& STEREO&        -- &this paper\\
F&2455028.8--2455051.0&            491& STEREO&        -- &this paper\\
F&2455247.6--2455266.8&            632& STEREO&        -- &this paper\\
F&2455417.7--2455436.1&            505& STEREO&        -- &this paper\\
F&2455592.1--2455610.4&            571& STEREO&        -- &this paper\\
F&2455813.1--2455829.4&            421& STEREO&        -- &this paper\\
F&2455939.4--2455955.7&            512& STEREO&        -- &this paper\\
F&2456200.7--2456217.1&            362& STEREO&        -- &this paper\\
G&2456523.6--2460583.5&            226& \ubvr &25330/26171&this paper\\
H&2456590.4--2456727.4&310/435/437/441& $BVRI$&      25330&this paper\\
I&2457732.2--2457746.9&           4314& MOST  &        -- &this paper\\
I&2458059.3--2458093.5&          17943& MOST  &        -- &this paper\\
J&2458438.1--2458463.7& 1214*)& TESS  &        -- &this paper\\
K&2458807.1--2459192.3&           2426&$RGB$  &    all-sky&\citet{yin2025} and priv.com.\\
J&2459174.2--2459200.2& 1248*)& TESS  &        -- &this paper\\
J&2459447.7--2459524.4& 3494*)& TESS  &        -- &this paper\\
J&2460208.6--2460259.9& 2422*)& TESS  &        -- &this paper\\
\noalign{\smallskip}\hline
\end{tabular}
\end{flushleft}
    \tablefoot{*)Normal data points over 0\fd02 to reduce
    the excessive number of observations}
\end{table*}

We carefully compiled all available published photometric observations with
known times of observations and combined them with new \ubvr\ photoelectric
observations secured at the Hvar Observatory and with $BVRI$ observations
secured in the private observatory of Robert Uhl\'a\v{r}. The log of all available 
photometric observations is in Table~\ref{jouphot}, 
and the details on photometric datasets
and their homogenization and reductions can be found in Appendix~\ref{apb}.
Inspecting the phase plots, we found that the data from files A, F, and K
have higher errors than all remaining datasets, and we excluded them from our
analyses. However, as shown in Fig.~\ref{excluded}, even
these data match the accurate solution found in the next section.

\section{\phoebe solution for all good photometric observations}
\begin{figure}
\resizebox{\hsize}{!}{\includegraphics{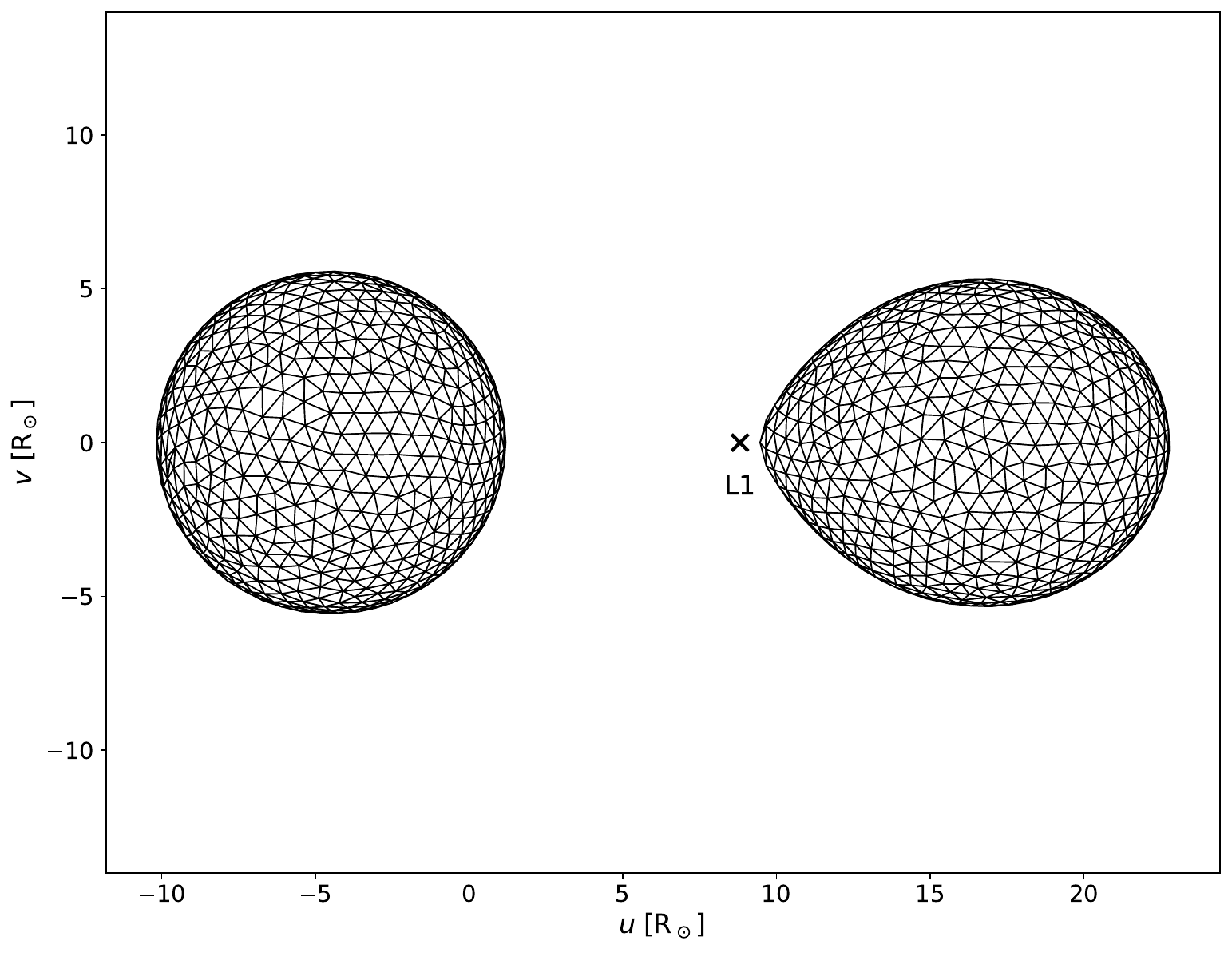}}
\resizebox{\hsize}{!}{\includegraphics{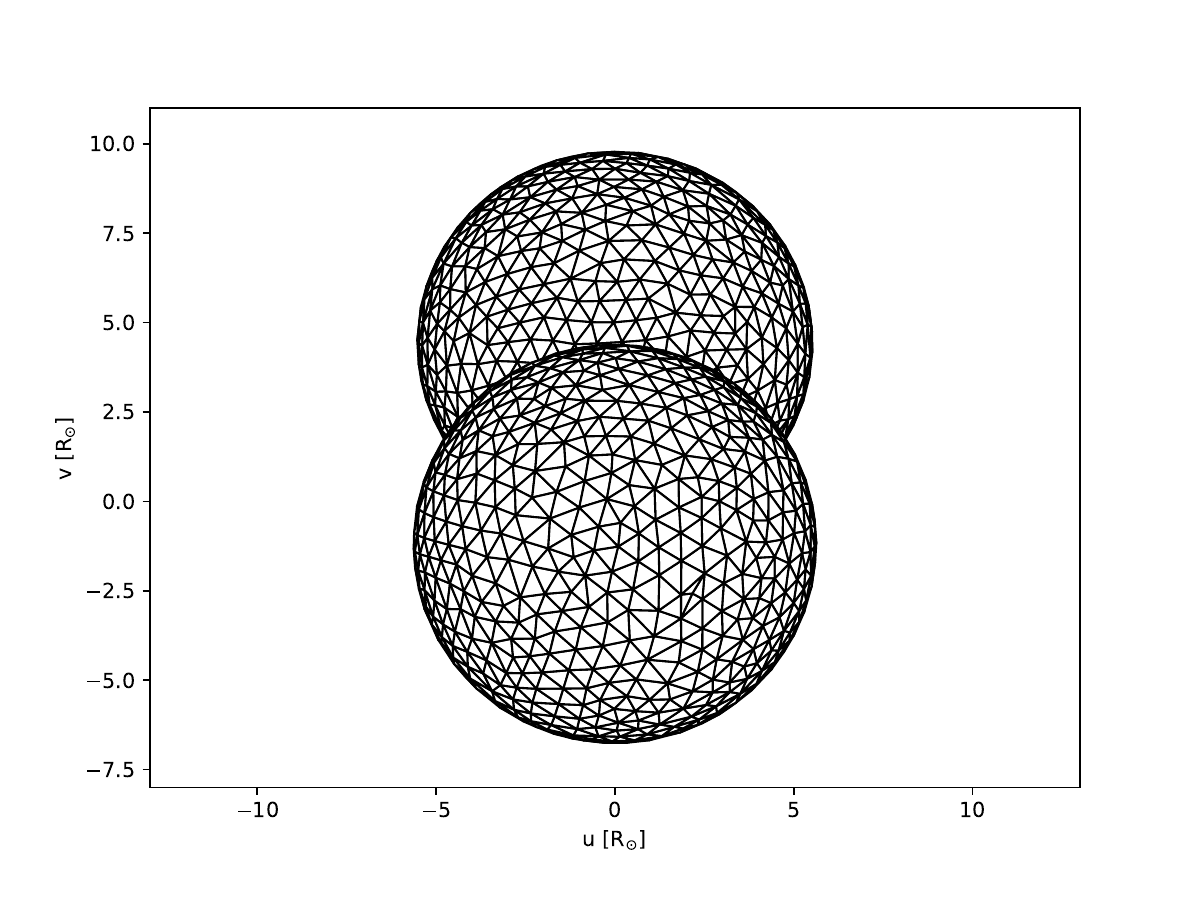}}
\caption{\textit{Top}: Mesh plot showing the eclipsing binary system at
elongation with the primary approaching. The position of the Lagrangian point
$L_1$ is shown.
\textit{Bottom}: Mesh plot at the phase of a secondary mid-eclipse.}\label{meshes}
\end{figure}

Since we succeeded in accumulating and homogenizing a very representative
collection of photometric observations covering an interval of 39100 days,
we first decided to verify the stability of the orbital period of the inner
system. To do this, we used
the programme \phoebe1 \citep{prsa2005, prsa2006}, which works quickly and is
quite sufficient for the given purpose. The inner binary was reported to be
a~semi-detached system, but as we mention later, there is a suspicion that
the mass transfer process can be non-continuous and that the Roche-lobe overflow
occurs only near the periastron passages of the outer system. For this reason,
we treated the binary as `detached' during our solution. Nevertheless, it did
turn out that the secondary is in or very close to the state of Roche lobe
filling. This is illustrated by the light curve model mesh plots shown in
Fig.~\ref{meshes} and by the fact that the side radius of the secondary
from the \phoebe solution corresponds to that of the Roche lobe for the obtained mass ratio.

\begin{figure}
\resizebox{0.9\hsize}{!}{\includegraphics{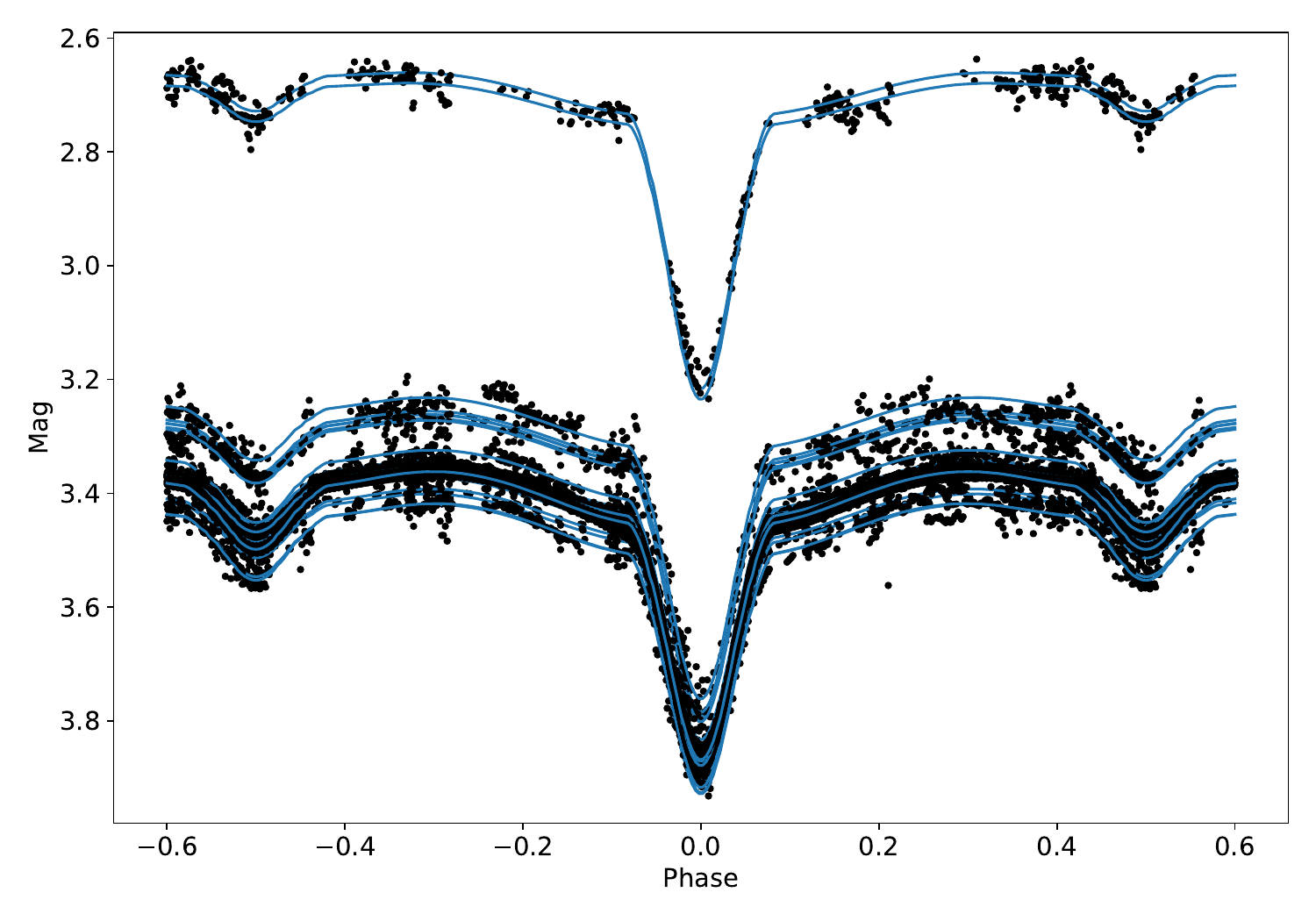}}
\resizebox{0.9\hsize}{!}{\includegraphics{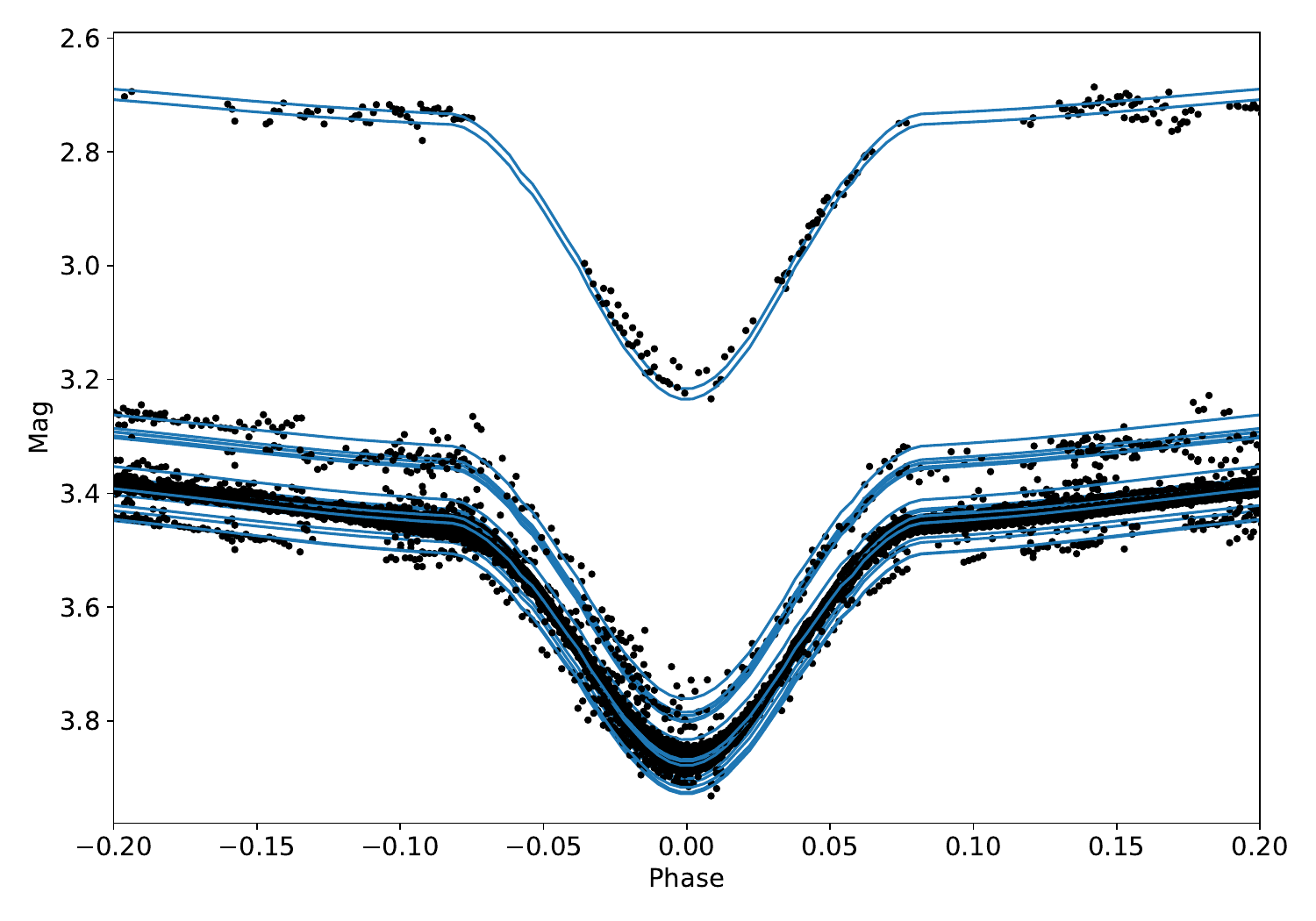}}
\caption{\textit{Top}: Light curves based on all available good photometries
for the final \phoebe solution.
\textit{Bottom}: Interval of phases near the primary minimum.}\label{lightcur}
\end{figure}

During modelling, we kept the effective temperature of the primary fixed
at 23000~K, which corresponds to its spectrum.
\footnote{We admit that this particular choice is somewhat arbitrary, but we have
verified that the period determination, which was our primary goal at this stage,
is robust even for reasonably different values of the effective temperature.}
Limb darkening coefficients
were used from the built-in PHOEBE models. We obtained an effective temperature
for the secondary of 9570~K; very similar relative side radii of 0.266 and 0.269
for the primary and secondary, respectively; and an orbital inclination of
$74.\!\!^\circ701\pm0.\!\!^\circ004$.
\footnote{The small uncertainty is the formal rms value of the \phoebe1 solution.}
We note that \citet{cester1978}
obtained a value of $76.\!\!^\circ0\pm0.\!\!^\circ5$, which they note as the mean
of the $V$ and $B$ values
($76.\!\!^\circ2\pm0.\!\!^\circ5$ and $75.\!\!^\circ4\pm0.\!\!^\circ6$).
\citet{berdy2018} found values from $75.\!\!^\circ1$ to $76.\!\!^\circ3$
from polarimetry.
The relative luminosities in the blue $B$ band were found to be 0.84 and 0.16
for the primary and secondary, respectively. These values are less extreme than the values of
0.92 and 0.08 estimated by \citet{fekel82}. The rotation of the secondary was
assumed to be synchronised with the binary orbital period, while the rotation rate
of the primary was assumed to be 1.4 times faster (based on the observed \vsin values).

Our solution indicates that the orbital period is very stable during the whole
time interval of available accurate observations, as seen in the phase plots
of all available light curves in Fig.~\ref{lightcur}. The corresponding linear
ephemeris is
\begin{equation}
T_{\rm min.I}={\rm HJD}\,2440998.86069(35)+3\fd952952845(81)\cdot E\,.\label{efe2}
\end{equation}
\noindent The secular constancy of the orbital period can also be illustrated
quantitatively by noting that the new ephemeris (\ref{efe2}) predicts
that the epoch of the primary minimum derived by \citet{steb20} is
RJD~21506.85021, which is in perfect agreement with the ephemeris (\ref{efe1}). We also
note that the very first minimum was observed at Hamburg on January 21, 1844,
at 7.334 hours UT. If we assume that the local Hamburg time of 8 hours was
the one used by \citet{schmidt1858}, namely HJD~2394587.3082, the ephemeris
(\ref{efe2}) would situate it at the orbital phase 0.017. Finally -- as
illustrated by Fig.~\ref{excluded} -- the less accurate STEREO and
Tibet Ali Observatory photometric observations are also fitted
very well by our new ephemeris. We therefore conclude that the orbital period
of the inner binary has been constant over the past 180 years of observations
(within the uncertainties stated in Eq.~\ref{efe2}).

\section{Preliminary orbital solutions}
In order to make use of all previously published RVs, we used the programme
\respefoe2 \citep[see the appendix in][]{hec2026} to determine the RVs from our new
spectra, and we combined the two sets of RVs into a single dataset.
Our RV values represent the mean values measured using \ion{He}{i} and
metallic lines. The Balmer lines were used to derive RVs for the primary
component only. This is because the position of the secondary line is always
strongly biased by being blended with the wide and steep absorption wings of
the primary line.

\begin{figure}
\resizebox{\hsize}{!}{\includegraphics{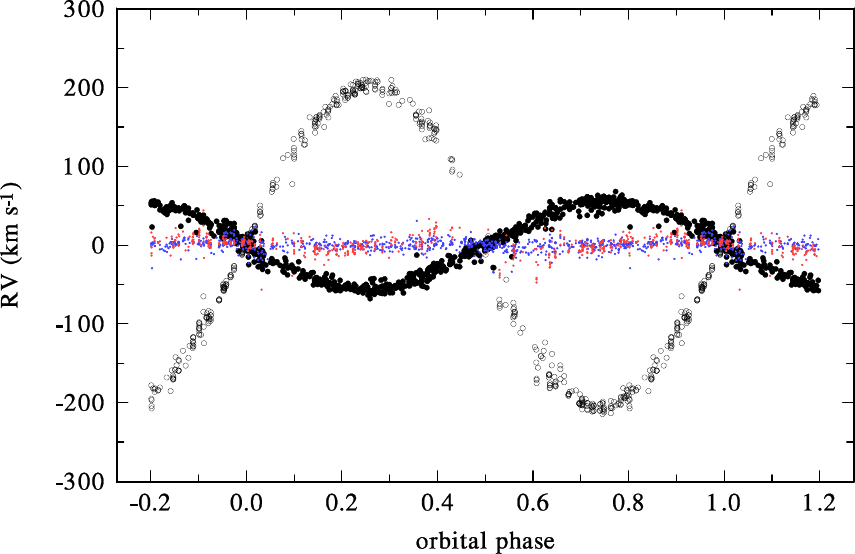}}
\caption{Preliminary RV curves based on mean RVs from the literature and
on ours measured in \respefoe2. In the process of the triple-star solution
using the \fotel program, the RVs were pre-whitened for the 33\fd025 orbit.
Black circles denote the primary RVs, open circles
the secondary RVs. The residuals from the global fit are shown by blue dots
for the primary, and red dots are for the secondary.}\label{inner}
\end{figure}

\begin{figure}
\resizebox{\hsize}{!}{\includegraphics{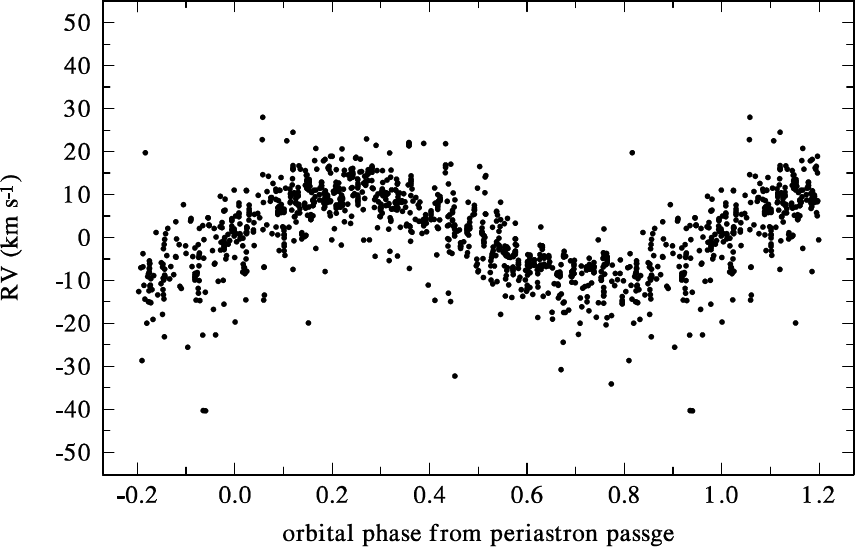}}
\resizebox{\hsize}{!}{\includegraphics{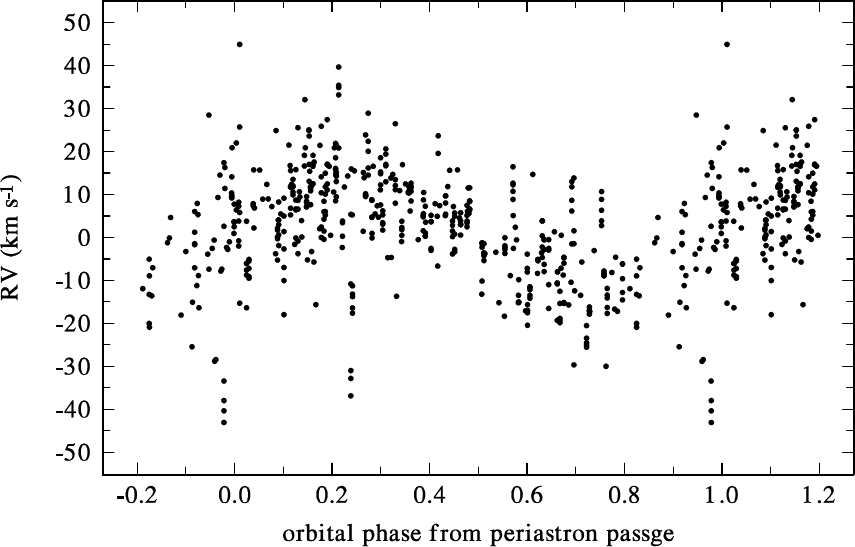}}
\caption{Preliminary average RV curve of the motion of the close pair
in the 33\fd025 orbit with the unseen tertiary, pre-whitened for the
motion in the inner orbit and based on the mean RVs.
\textit{Top}: RV curve based on the residuals of the primary RVs
for the short orbit. \textit{Bottom}: Noisier RV curve
based on the residuals of the secondary RVs.}\label{outer}
\end{figure}

Next, we used the programme \fotel \citep{fotel1,fotel2} to obtain the preliminary
orbital solutions. We found this to be an adequate initial step since \fotel provides a
simultaneous solution for the inner and outer orbits. Nevertheless, there are two
simplifying assumptions we made: (i) We fixed the eccentricity of the inner orbit
to zero, and (ii) we considered the orbital parameters of the inner and outer orbits to be
constant (see Sect.~\ref{orbi} for more discussion). We also kept the period
and the epoch of the primary mid-eclipse fixed at the
values derived with \phoebe1. After several trial solutions,
demonstrating good agreement between separate solutions based on either
only primary or only secondary RVs, we determined the final solution based
on all primary and secondary RVs. To infer possible
systematic effects of the measurements and to estimate rms errors separately
for the primary and secondary RVs, respectively, an independent value of the
systemic velocity was used for the two sets of RVs. The orbital elements are
provided in Table~\ref{fotel}, and the corresponding RV curves are shown
in Figs.~\ref{inner} and \ref{outer}, each folded to the phase of
the corresponding orbital period.

\begin{table}[t]
\begin{flushleft}
\caption{Orbital elements based on all published RVs and on mean \respefoe2 RVs
        for both binary components of the eclipsing subsystem.}
\label{fotel}
\begin{tabular}{lccrrccc}
\hline\hline\noalign{\smallskip}
        Element                   & Value             \\
\noalign{\smallskip}\hline\noalign{\smallskip}
$P$ (d)                    & 3.952952845 fixed        \\
$T_{\rm Min.1}$ (RJD)      &  40998.86069 fixed       \\
$e$                        &   0.0 fixed              \\
$\gamma_1$ (\ks)           &$17.08\pm0.20$            \\
$\gamma_2$ (\ks)           &$16.48\pm0.50$            \\
$K_1$    (\ks)             & $55.51\pm0.28$           \\
$K_1/K_2$                  &$0.2712\pm0.0017$         \\
$K_2$    (\ks)             & 205.02                   \\
\noalign{\smallskip}\hline\noalign{\smallskip}
$P_{\rm long}$ (d)         & $33.02480\pm0.00028$     \\
$T_{\rm periast.}$ (RJD)   & $56822.5\pm3.6$          \\
$T_{\rm RV max.}$ (RJD)    & 56830.3                  \\
$e$                        & $0.054\pm0.033$          \\
$\omega$ ($^\circ$)        & $269\pm39$               \\
$K_{1+2}/K_1$              & $0.1927\pm0.0060$        \\
$K_{1+2}$ (\ks)            & 10.70                    \\
$m_1\sin^3i$ (\ms)         & 5.677                    \\
$m_2\sin^3i$ (\ms)         & 1.540                    \\
$a\,\sin i$ (\rs)          & 20.33                    \\
No. of RVs (prim./sec.)    & 993/499                  \\
rms${_1}$ (\ks)            &  6.14                    \\
rms${_2}$ (\ks)            & 10.95                    \\
\hline\noalign{\smallskip}
\end{tabular}
\end{flushleft}
\tablefoot{Separate systemic velocities were allowed for the primary and secondary.
A triple-star solution was derived with the program \fotele.}
\end{table}

\section{Spectra disentangling with \korele}
\begin{figure}
  \resizebox{\hsize}{!}{\includegraphics{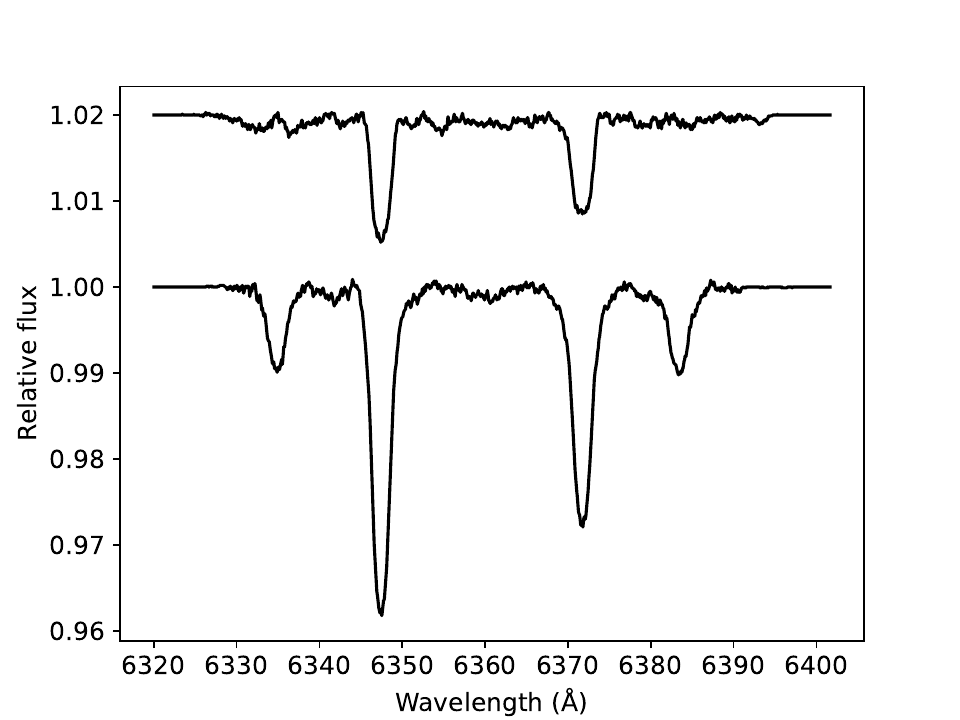}}
  \resizebox{\hsize}{!}{\includegraphics{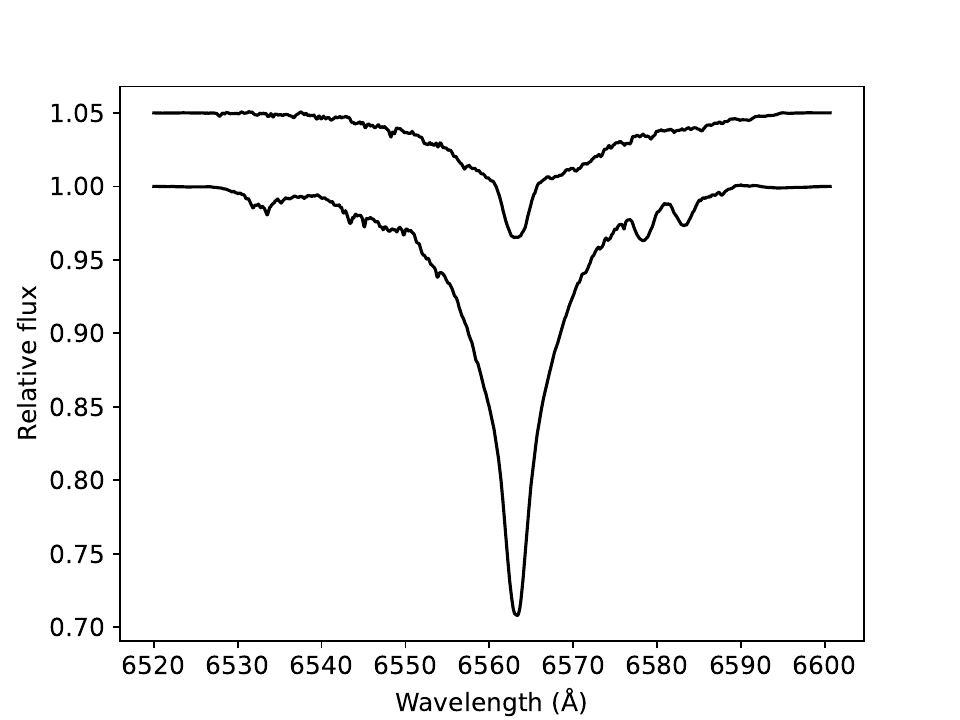}}
  \resizebox{\hsize}{!}{\includegraphics{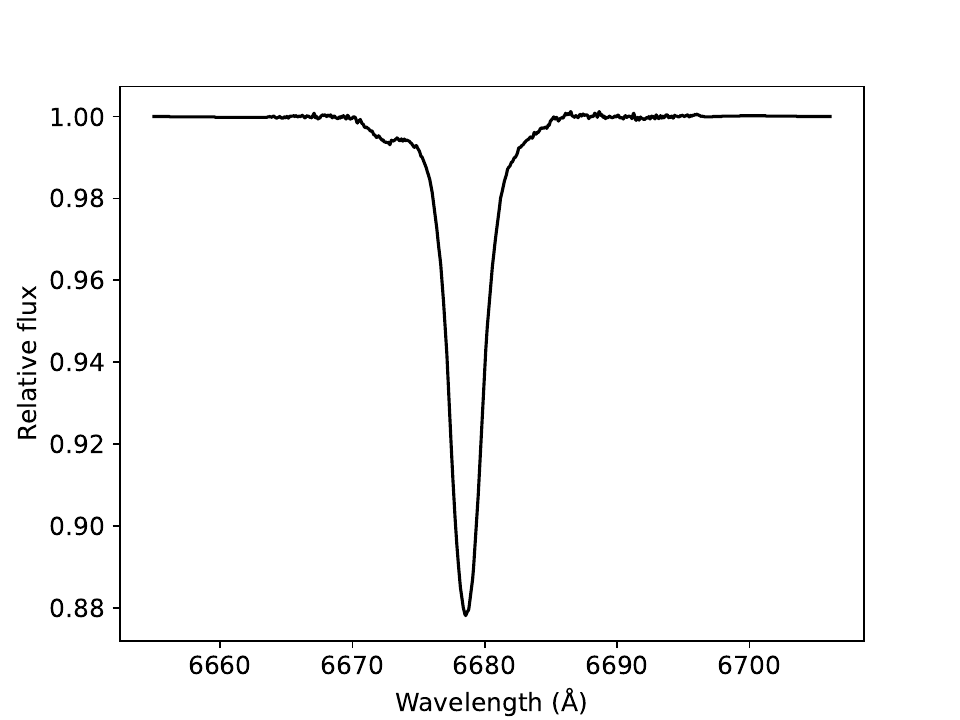}}
  \caption{Normalised disentangled line profiles.
Bottom profiles correspond to the primary component. Those from the secondary have been
shifted upwards for a constant in the relative flux.}
  \label{korel_lines}
\end{figure}

To verify the results of standard RV measurements, we analysed all available
electronic spectra with a more sophisticated approach, namely spectra disentangling.
First, we derived trial solutions in which the spectra were split into several subsets according to their time of observations. Individual subsets covered time intervals no longer than 1 year.
We separately disentangled every subset that contained at least ten spectra using
the programme \korel \citep{korel1,korel2,korel3}. All trials and final calculations were carried out by Jana \v{S}vr\v{c}kov\'a with the help of the VO-KOREL service (see
\citet{vokorel}). Since the results were consistent across
seasons and different spectral lines, in the end we disentangled all available
spectra at once in spectral segments centred on the \ion{Si}{ii}~6347 and 6371~\AA\
doublet, on the \ha line, and on \ion{He}{i}~6678~\AA\ (see Fig.~\ref{korel_lines})
to obtain the solutions given in Table~\ref{korel}. The practical reason for this
choice was that these spectral segments were available for almost all spectra we
used. Not surprisingly, there is no trace of the secondary star in the \ion{He}{i}
line, while both components were disentangled in the \ha and \ion{Si}{ii} lines.

The \korel solution of Table~\ref{korel} agrees remarkably well with the overall \fotel solution
of Table~\ref{fotel} and also with the previous result obtained by \citet{fekel82}.
Nevertheless, we point out two issues of tension in the two
results. First, the solution of \fotel leads to a considerably lower
eccentricity of the outer orbit than \korel. Second, there appears to be no
apsidal precession of the outer orbit over the 18 years spanned by the
spectral observations in the \korel solution (even though the software
would provide a linear drift in this orbital parameter if it
resulted in a better fit of the data; see also Sect.~\ref{orbi}).
We believe the two problems are related and will
be resolved by using a full-fledged dynamical model of the system in the planned
forthcoming study. On the positive side, we note that the epochs
of the periastron passage of the outer orbit agree well for \fotele~and
all \korel solutions.

\begin{table}[t]
\caption{\label{korel} \korel solutions for the three spectral segments and a solution (`red') based on all three spectral segments combined.}
\begin{tabular}{l c c c c}
\hline\hline \noalign{\smallskip}
Element & \hae &  Si II & He 6678 & red \\   \hline \noalign{\smallskip}
$P$ (d) & \multicolumn{4}{c}{3.952952845 -- fixed} \\
$T_\mathrm{Min.1}$ (RJD) & \multicolumn{4}{c}{40998.86069 -- fixed} \\
$e$ & \multicolumn{4}{c}{0.0 -- fixed} \\
$K_1$ (\ks) & 57.87 &   56.28   &   59.25 & 56.59\\
$q = K_1/K_2$ & 0.278   &   0.276   &   --- & 0.273\\
$K_2$ (\ks) &   208.253 &   203.670 &   --- & 207.273\\
\noalign{\smallskip} \hline \noalign{\smallskip}
$P_\mathrm{long}$ (d) & 33.02738&33.02750   &33.01802 & 33.02975 \\
$T_\mathrm{periastr.}$ (RJD) & 56822.62 &56821.89 &56822.13 & 56821.78 \\
$e$ & 0.200 &   0.149   &   0.102 & 0.154 \\
$\omega$ ($^\circ$) & 274.2 &   266.6   &   265.4 & 269.22 \\
$K_{1+2}$ (\ks) & 9.71  &   10.51   &   10.63 & 9.95\\
\noalign{\smallskip} \hline
\end{tabular}
\end{table}

\section{Preliminary system properties}

By adopting the \korel fit to the whole set of available spectra as a nominal
solution, we obtained
the following values for the inner subsystem: $m_1\sin^3i=5.911$~\ms,
$m_2\sin^3i=1.614$~\ms, and  $a\sin i=20.616$~\rs. For the inclination
of $74.\!\!^\circ7$ and the relative side radii of the \phoebe1 solution,
this implies quite plausible values of $m_1=6.59$~\ms and $m_2=1.80$~\ms,
a semi-major axis of $a=21.374$~\rs, and component radii of $R_1=5.69$~\rs and
$R_2=5.75$~\rs. In contrast to the result of \citet{fekel82} we found
that the Roche-lobe secondary is slightly larger than the primary.
For the outer subsystem, adopting again the global \korel solution
with $K_{1+2}=9.95$~\ks, $e=0.154$, the same system
inclination, and the total mass of the inner binary of 8.39~\ms,
we obtained $m_3=0.667$~\ms, $a_{\rm outer}=90.09$~\rs, and a separation
at periastron of $a_{\rm outer}(1-e)=76.22$~\rs. The mass of the third body
would correspond to a K4 star, according to the tabulation by \citet{mr}.

Returning to the remarkable stability of the eclipsing
binary period ($P$) for over more than a century (Eq.~\ref{efe2}), we
can estimate an upper limit for the hypothetical mass
transfer from the secondary to the primary. Assuming for
simplicity an adiabatic transfer conserving total mass
and angular momentum of the system, the rate of mass
loss from the secondary produces a period change whose
values are related by
\begin{equation}
 \frac{d\mu}{dt} = \frac{1}{3}\frac{\mu\left(1-\mu\right)}{2\mu-1}
  \, \frac{1}{P}\frac{dP}{dt}, \label{massloss}
\end{equation}
where $\mu=m_2/(m_1+m_2)$
\citep[see, for example, Eq. 31 in][]{sepin2007}.
Having found that $P=P_0\pm \Delta P_0$ (orbital period and its rms error)
over a~time interval $T$, the upper limit of the
period change is $dP/dt< 2\,\Delta P_0/T$. Using this
estimate in Eq. (\ref{massloss}), we obtained the following:
\begin{equation}
 \frac{d m_2}{dt} < \frac{2}{3}\frac{\mu\left(1-\mu\right)}{2\mu-1}
  \, \frac{\Delta P_0}{P_0}\,\frac{m_1+m_2}{T}. \label{massloss1}
\end{equation}
Adopting the values $P_0$ and $\Delta P_0$ from Eq.~(2), $T\simeq 100$~yr
and the total mass $m_1+m_2\simeq 8.39$~M$_\odot$ estimated above, we
have $|d\,m_2/dt|< 3.3\times 10^{-10}$~M$_\odot$~yr$^{-1}$. This is a very low
but reasonable value. We note that in semi-detached systems with the detected
change in orbital period, the rates
of mass transfer are usually of the order of $10^{-9}$~M$_\odot$~yr$^{-1}$ or
higher \citep[see, e.g.,][]{celik2026}. Despite a non-detection of the
orbital period change for \lae, the upper limit on
magnitude of the mass-transfer rate may be useful if this process had to
be connected with photometric micro-variability
\citep[see Sect.~\ref{micro} and][]{berdy2018}.
With the orbital solution caveats discussed above, which could also influence 
the inferred values of physical parameters to some degree, we consider the latter
to be a valuable initial estimate for more in-depth modelling in the future.

\section{Some peculiar phenomena worth of further investigation}
In this section we briefly highlight a couple of puzzling phenomena that
we noticed in our analysis. We believe that they are just different
manifestations of the fact that \lae ~is a highly dynamical system down 
to the shortest timescale of the orbital period of the eclipsing binary. 

\subsection{Occasional line-core asymmetries or even line doubling 
for the primary component}
\begin{figure}
\resizebox{\hsize}{!}{\includegraphics{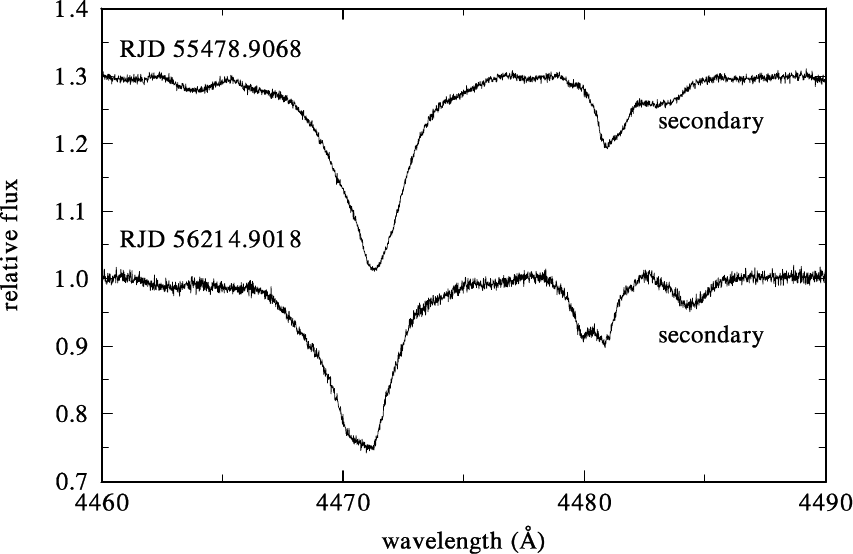}}
\caption{Two peculiar HARPS spectra in the neighbourhood of the
\ion{He}{i}~4471~\AA\ and \ion{Mg}{ii}~4481~\AA\ lines.}\label{harps}
\end{figure}

While measuring the RVs interactively in \respefoe2, we noted that especially
the \ion{He}{i} lines and the \ion{Mg}{ii}~4481~\AA\ line of the primary
component exhibit either blue or red asymmetries in the line cores 
and even the line doubling. Two outstanding examples of such variations are
shown in Fig.~\ref{harps}. They do not seem to correlate with any
particular orbital phase of the close eclipsing subsystem, and they were observed
far from the epochs of the periastron passage in the outer orbit. They
might indicate some temporal changes in the large-scale mass transfer
between the binary components.

\subsection{Physical microvariability of the brightness}\label{micro}
\begin{figure}
\resizebox{\hsize}{!}{\includegraphics{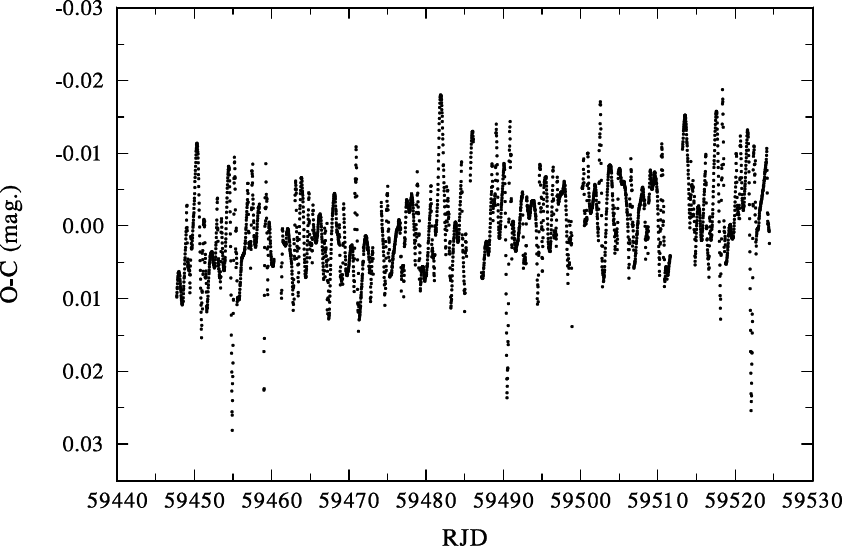}}
\caption{\oc\ residuals from the \phoebe1 light curve fit of
accurate TESS photometry from segments 42, 43, and 44 versus time.
}\label{oc4244}
\end{figure}

During inspection of the TESS and MOST light curves, we observed that some
seemingly complicated microvariability is present. As an example, we show
the \oc\ residuals of the \phoebe1 light curve fit for the long series
of TESS observations in segments 42, 43, and 44 in Fig.~\ref{oc4244}.

In this context, we note that \citet{berdy2018} analysed non-periodic phenomena
in their polarisation observations, which they attributed to sporadic changes
in the mass transfer rate from the secondary to the primary of the eclipsing
subsystem. It is tempting to connect them with the photometric microvariability
shown in Fig.~\ref{oc4244}. The limit on the mass-transfer rate
derived above may be a useful quantitative constraint for further modelling.

\subsection{Eclipse timing variations}
The very accurate photometry provided by the \textit{Kepler} and TESS surveys and their
discoveries of very compact triples and higher stellar hierarchies enabled analysis 
and scientific use of eclipse timing variations 
\citep[ETVs; see, e.g.,][]{borko2022}.
Basically, ETVs are differences between the epochs of the
primary and secondary light curve minima predicted by the linear ephemeris
(\ref{efe2}) and their true instants. In compact systems such as \lae, ETVs
are primarily caused by gravitational interaction between the inner and outer orbits
(often referred to as physical ETVs) overshadowing the traditional light travel
time effect \citep{borko2022}.

Although the ETV analysis proved very useful in the studies of many systems,
here a~complication stemed from the micro-variability of the photometric
signal well documented away from the eclipse epoch (Sect.~\ref{micro}).
This caused a perturbation propagating towards the eclipses, making
the photometric signal near the minima asymmetric and time variable.
The inferred ETV series may thus contain a non-negligible noise
component in the case of \lae.

\begin{figure}
 \resizebox{\hsize}{!}{\includegraphics{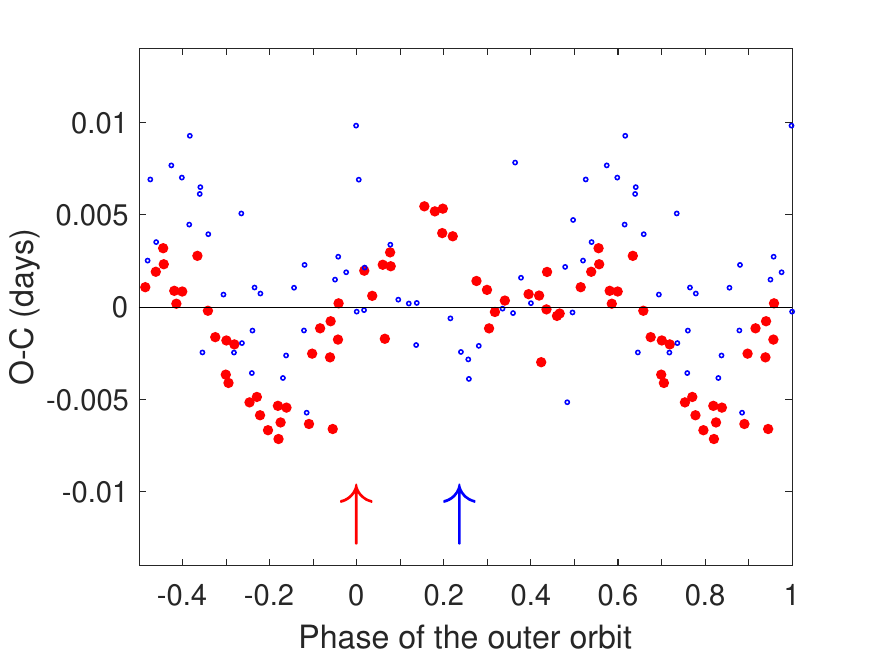}}
 \caption{Eclipse timing variation diagram of the times of
the mid-eclipse of the inner orbit, plotted versus phase of the outer
orbit from Table~\ref{fotel}. Larger red dots denote the primary, and
smaller blue dots indicate the secondary eclipses. The red arrow indicates the
instant of the periastron passage, while the blue
one represents the time of maximum RV.}
 \label{ETVdiagram}
\end{figure}

With this caveat in mind, we gave a preliminary look at the ETVs derived
from all available TESS and MOST data (seven sectors of TESS and two seasons of
MOST; Table~3). We used three different methods
(fitting the light curve template, the classical Kwee-van Woerden method,
and the polynomial fitting method) to derive mid-eclipse epochs for all
primary and secondary eclipses. The final ETV data, representing
the robust mean of the three individual values, are shown in Fig.~\ref{ETVdiagram}.
Since the principal physical ETV signal has the period of the outer
orbit in the system, we used the phase of the outer orbit at the abscissa
instead of time in Fig.~\ref{ETVdiagram}. The fundamental quadrupole
interaction ETV signal has a~predicted amplitude
 \citep[see, e.g., Eqs.~(20) and (5) in][respectively]{borko2011,borko2015}:
\begin{equation}
 \delta t \simeq \frac{3e}{2\pi} \frac{m_3}{m_1+m_2+m_3}\frac{P^2}{P_{\rm
  long}}\simeq 2.5\times 10^{-3}\;\rm{d}\, ,
\end{equation}
\noindent where we used $e\simeq 0.15$ for the outer orbit eccentricity. However,
additional contributions will certainly arise due to a significant
octupole orbital interaction (due to non-equal masses $m_1$ and $m_2$ of
the binary) and apsidal and tidal contributions. In particular, the apsidal
contribution depends sensitively on the mean eccentricity of the inner
orbit. In this paper, we fixed its value to zero, but in Sect.~\ref{orbi}
we argue that a small mean value of up to $e_{\rm in}\simeq 0.005$ should be forced
by the gravitational perturbation from the third component in the system.
If so, the apsidal contribution to ETVs is of the order of $\delta t\simeq
Pe_{\rm in}/\pi$, a few thousand a day \citep[e.g.,][]{gb1995}.
At this moment we can conclude
that the signal shown in Fig.~\ref{ETVdiagram} has the expected order
of magnitude, has the principal phasing with the period of the outer orbit, and
(most importantly) is not dominated by noise. We plan to make full use of
these ETV series in our forthcoming study.

\subsection{Puzzles of the preliminary orbital solution of $\lambda$~Tau}
\label{orbi}
Given the compactness of the hierarchical triple system $\lambda$~Tau, we
 expect it to evolve highly dynamically \citep[see, e.g.][]{borko2022}.
In this respect, the absence of apsidal motion of the outer orbit
appears to be a curious element of our results (Sect.~6). Indeed,
adopting a~near-coplanar geometry of the system, a simple estimate
of the outer orbit apsidal rate from the quadrupole interaction of the
inner and outer orbits would be
\begin{equation}
 \frac{d\omega}{dt}\simeq n_{\rm long}\,\frac{3}{4}\frac{m_1m_2}{\left(m_1+m_2\right)^2}
   \left(\frac{m_1+m_2}{m_1+m_2+m_3}\right)^{2/3}\left(\frac{P}{P_{\rm long}}
   \right)^{4/3}, \label{aps2}
\end{equation}
where $n_{\rm long}=2\pi/P_{\rm long}$ is the mean motion of the outer orbit;
see, for example, Eq.~(36) in \citet{sordel75} or
Eq.~(86) in \citet{breiter2015}. With all the parameters on the right-hand side
of (\ref{aps2}) known, we would
expect $\omega$ to circulate with a period of $\sim 13.2$~years.
A direct numerical simulation, which also contains higher-order orbital
couplings, confirms this value within an uncertainty of $\sim 20$\%.
Either our analysis suffers from oversimplifications that propagate principally
to the solution of outer orbit periapse behaviour or some
unexpected dynamical processes have to be included in the future solution.

As far as the first possibility is concerned, the main issue may be related to
our assumption of zero eccentricity, $e_{\rm in}$, of the inner orbit.
However, the significant difference in masses $m_1$ and $m_2$ triggers
an octupole interaction between the inner and outer orbits, which should force
a~small but non-zero binary eccentricity. Our numerical tests as well as
previous analytical estimates \citep[e.g.][]{georg2003,georg2009}
provide a mean value of $e_{\rm in}\sim 0.005$ for the inner eccentricity.
This value is small and may not be easily detectable in the available
photometry or spectroscopic observations directly.
However, a cumulated effect of the inner orbit apsidal precession could
provide more hints from the analysis of the ETVs of the eclipsing binary.
Our preliminary data shown in Fig.~\ref{ETVdiagram} still leave room for
apsidal contribution, but they need to be analysed using a more complex approach.

\section*{Data availability}
Tables A.1 and B.1 -- B.8 are available only in electronic form
at the CDS via anonymous ftp to cdarc.u-strasbg.fr (130.79.128.5)
or via http://cdsweb.u-strasbg.fr/cgi-bin/qcat?J/A+A/ .

\begin{acknowledgements}
We acknowledge the use of the programme \phoebe developed by Andrej Pr\v{s}a,
the programmes \fotel and \korele, written by Petr Hadrava, and the programme \respefoe2
by Adam Harmanec. Some of the Ond\v{r}ejov spectra were obtained by
Jana Nemravov\'a - Bene\v{s}ov\'a (who left astronomical research in the meantime)
and also by the late Pavel Mayer. The authors also thank Jia Yin, who kindly
provided us with his all-sky RGB photometry from Ali Observatory in Tibet,
and Danielle Bewster for advice on STEREO photometry.
Based on new spectroscopic observations
from the following observatories: Cerro Tololo Inter-American Observatory,
ESO LaSilla, Ond\v{r}ejov, Star\'a Lesn\'a, and Lisboa, and on new
photometry from Hvar, Hipparcos, STEREO, and Ali Observatory, and also
on the very last photometric observations carried out in 2016 and 2017 by
the memorable Canadian space observatory MOST and downloaded
through the Vienna tracking station, often with a heroic effort by the staff
there. The paper also includes data collected by the TESS mission, which are
publicly available from the Mikulski Archive for Space Telescopes (MAST).
Funding for the TESS mission is provided by NASA's Science Mission directorate.
Over the years, the research of the Czech authors was supported by the grants
205/06/0304, P209/10/0715, and GA15-02112S of the Czech
Science Foundation, 678212 and 250015 of the Charles University Grant Agency,
research project AV0Z10030501 of the Academy of Sciences of the Czech Republic,
Research Programme MSM0021620860 Physical study of objects and processes
in the solar system and in astrophysics of the Ministry of Education of
the Czech Republic.
Hrvoje Bo\v{z}i\'c obtained financial support from the Croatian Science Foundation
under the project 6212 ``Solar and Stellar Variability". The research of Theo
Pribulla was supported by the VEGA grant of the Slovak Academy of Sciences
No. 2/0033/26 and by the Slovak Research and Development Agency under contracts
Nos. APVV-24-0160 and APVV-25-0583. Constructive suggestions from
an anonymous referee helped improve the clarity of the paper.
The authors acknowledge the use of the electronic
bibliography maintained by the NASA/ADS system and by the CDS in Strasbourg.
\end{acknowledgements}

\bibliographystyle{aa}
\bibliography{lamtau}

@ARTICLE{bax1848,
   author = {{Baxendell}, J.},
    title = "{On the variability of {$\lambda$} Tauri}",
  journal = {\mnras},
     year = 1848,
    month = dec,
   volume = 9,
    pages = {37},
   adsurl = {http://adsabs.harvard.edu/abs/1848MNRAS...9...37B}
}

@ARTICLE{belo1898,
   author = {{Belopolsky}, A.},
    title = "{{\"U}ber das Spectrum von {$\lambda$} Tauri (3$^{m}$.4 - 4$^{m}$.2)}",
  journal = {Astronomische Nachrichten},
     year = 1898,
    month = feb,
   volume = 145,
    pages = {281},
      doi = {10.1002/asna.18981451803},
   adsurl = {http://adsabs.harvard.edu/abs/1898AN....145..281B}
}

@ARTICLE{berdy2018,
       author = {{Berdyugin}, A. and {Piirola}, V. and {Sakanoi}, T. and {Kagitani}, M. and {Yoneda}, M.},
        title = "{High-precision broad-band linear polarimetry of early-type binaries.
		II. Variable, phase-locked polarization in triple Algol-type system {\ensuremath{\lambda}} Tauri}",
      journal = {\aap},
         year = 2018,
        month = mar,
       volume = {611},
          eid = {A69},
        pages = {A69},
          doi = {10.1051/0004-6361/201732163},
       adsurl = {https://ui.adsabs.harvard.edu/abs/2018A&A...611A..69B}
}

@ARTICLE{bewsher2010,
       author = {{Bewsher}, D. and {Brown}, D.~S. and {Eyles}, C.~J. and {Kellett}, B.~J. and {White}, G.~J. and {Swinyard}, B.},
        title = "{Determination of the Photometric Calibration and Large-Scale Flatfield of the STEREO Heliospheric Imagers: I. HI-1}",
      journal = {\solphys},
         year = 2010,
        month = jul,
       volume = {264},
       number = {2},
        pages = {433-460},
          doi = {10.1007/s11207-010-9582-8},
       adsurl = {https://ui.adsabs.harvard.edu/abs/2010SoPh..264..433B}
}

@INPROCEEDINGS{bolt2007,
   author = {{Bolton}, C.~T. and {Grunhut}, J.~H.},
    title = "{The Orbit and Properties of the Spectroscopic-Eclipsing- Interferometric Trips System {$\xi$} Tauri}",
booktitle = {IAU Symposium},
     year = 2007,
   series = {IAU Symposium},
   volume = 240,
   editor = {{Hartkopf}, W.~I. and {Harmanec}, P. and {Guinan}, E.~F.},
    month = aug,
    pages = {66},
   adsurl = {http://adsabs.harvard.edu/abs/2007IAUS..240...66B}
}

@ARTICLE{borko2011,
       author = {{Borkovits}, T. and {Csizmadia}, Sz. and {Forg{\'a}cs-Dajka}, E. 
	       and {Heged{\"u}s}, T.},
        title = "{Transit timing variations in eccentric hierarchical triple exoplanetary 
		systems. I. Perturbations on the time scale of the orbital period of 
		the perturber}",
      journal = {\aap},
         year = 2011,
        month = apr,
       volume = {528},
          eid = {A53},
        pages = {A53},
          doi = {10.1051/0004-6361/201015867},
archivePrefix = {arXiv},
       eprint = {1010.0884},
 primaryClass = {astro-ph.EP},
       adsurl = {https://ui.adsabs.harvard.edu/abs/2011A&A...528A..53B}
}

@ARTICLE{borko2015,
       author = {{Borkovits}, T. and {Rappaport}, S. and {Hajdu}, T. and {Sztakovics}, J.},
        title = "{Eclipse timing variation analyses of eccentric binaries with close 
		tertiaries in the Kepler field}",
      journal = {\mnras},
         year = 2015,
        month = mar,
       volume = {448},
       number = {1},
        pages = {946-993},
          doi = {10.1093/mnras/stv015},
archivePrefix = {arXiv},
       eprint = {1412.5759},
 primaryClass = {astro-ph.SR},
       adsurl = {https://ui.adsabs.harvard.edu/abs/2015MNRAS.448..946B}
}

@ARTICLE{borko2022,
       author = {{Borkovits}, Tam{\'a}s},
        title = "{Eclipsing Binaries in Dynamically Interacting Close, Multiple Systems}",
      journal = {Galaxies},
         year = 2022,
        month = jan,
       volume = {10},
       number = {1},
          eid = {9},
        pages = {9},
          doi = {10.3390/galaxies10010009},
archivePrefix = {arXiv},
       eprint = {2201.01243},
 primaryClass = {astro-ph.SR},
       adsurl = {https://ui.adsabs.harvard.edu/abs/2022Galax..10....9B}
}

@ARTICLE{borko2025,
       author = {{Borkovits}, T. and {Rappaport}, S.~A. and {Mitnyan}, T. and {B{\'\i}r{\'o}}, I.~B. and {Cs{\'a}nyi}, I.
	       and {Forg{\'a}cs-Dajka}, E. and {Forr{\'o}}, A. and {Hajdu}, T. and {Seli}, B. and {Sztakovics}, J.
	and {G{\"o}bly{\"o}s}, A. and {P{\'a}l}, A.},
        title = "{Then and now: A new look at the eclipse timing variations of hierarchical triple star candidates
		in the primordial Kepler field, revisited by TESS}",
      journal = {\aap},
         year = 2025,
        month = mar,
       volume = {695},
          eid = {A209},
        pages = {A209},
          doi = {10.1051/0004-6361/202453616},
archivePrefix = {arXiv},
       eprint = {2502.09480},
 primaryClass = {astro-ph.SR},
       adsurl = {https://ui.adsabs.harvard.edu/abs/2025A&A...695A.209B}
}

@ARTICLE{breiter2015,
       author = {{Breiter}, S. and {Vokrouhlick{\'y}}, D.},
        title = "{Secular motion in a hierarchic triple stellar system}",
      journal = {\mnras},
         year = 2015,
        month = may,
       volume = {449},
       number = {2},
        pages = {1691-1703},
          doi = {10.1093/mnras/stv361},
       adsurl = {https://ui.adsabs.harvard.edu/abs/2015MNRAS.449.1691B}
}

@ARTICLE{broz2026,
      author = {{Bro{\v{z}}}, M. and {Dole{\v{z}}al}, P. and {Vokrouhlick{\'y}}, D. and {Harmanec}, P. and {Barlow}, B. and {Bo{\v{z}}i{\'c}}, H. and {Labadie-Bartz}, J. and {Kuschnig}, R. and {Kallinger}, T. and {Matthews}, J. and et al.},
       title = "{Configuration of the {\ensuremath{\xi}} Tau system constrained by multi-technique observations}",
     journal = {\aap},
        year = 2026,
       month = jul,
      volume = {711},
         eid = {A297},
       pages = {A297},
         doi = {10.1051/0004-6361/202660047},
archivePrefix = {arXiv},
      eprint = {2605.22545},
primaryClass = {astro-ph.SR},
      adsurl = {https://ui.adsabs.harvard.edu/abs/2026A&A...711A.297B}
}

@ARTICLE{casi68,
   author = {{Casini}, C. and {Galeotti}, P. and {Guerrero}, G.},
    title = "{The Triple System Lambda Tauri}",
  journal = {IBVS No.},
     year = 1968,
    month = feb,
   volume = 257,
    pages = {1},
   adsurl = {http://adsabs.harvard.edu/abs/1968IBVS..257....1C}
}

@ARTICLE{celik2026,
       author = {{{\c{C}}elik}, E. and {Kahraman Ali{\c{c}}avu{\c{s}}}, F.},
        title = "{Investigating Potential Relationships Between Mass Transfer Indicators and $??$ Scuti Pulsations in oEA Systems}",
      journal = {arXiv e-prints},
         year = 2026,
        month = sep,
          eid = {arXiv:2609.08004},
        pages = {arXiv:2609.08004},
archivePrefix = {arXiv},
       eprint = {2609.08004},
 primaryClass = {astro-ph.SR},
       adsurl = {https://ui.adsabs.harvard.edu/abs/2026arXiv260908004C}
}

@ARTICLE{cester1978,
       author = {{Cester}, B. and {Fedel}, B. and {Giuricin}, G. and {Mardirossian}, F. and {Mezzetti}, M.},
        title = "{II. Revised photometric elements of the eclipsing binaries LY Aur, AS Eri, AW Peg, lambda Tau.}",
      journal = {\aap},
         year = 1978,
        month = jan,
       volume = {62},
        pages = {291-294},
       adsurl = {https://ui.adsabs.harvard.edu/abs/1978A&A....62..291C}
}

@ARTICLE{ebbig56,
   author = {{Ebbighausen}, E.~G. and {Struve}, O.},
    title = "{The Triple System Lambda Tauri.}",
  journal = {\apj},
     year = 1956,
    month = nov,
   volume = 124,
    pages = {507},
      doi = {10.1086/146254},
   adsurl = {http://adsabs.harvard.edu/abs/1956ApJ...124..507E}
}

@ARTICLE{fekel82,
   author = {{Fekel}, Jr., F.~C. and {Tomkin}, J.},
    title = "{Secondaries of eclipsing binaries. IV - The triple system Lambda Tauri}",
  journal = {\apj},
     year = 1982,
    month = dec,
   volume = 263,
    pages = {289-301},
      doi = {10.1086/160503},
   adsurl = {http://adsabs.harvard.edu/abs/1982ApJ...263..289F}
}

@ARTICLE{georg2003,
       author = {{Georgakarakos}, Nikolaos},
        title = "{Eccentricity evolution in hierarchical triple systems with eccentric outer binaries}",
      journal = {\mnras},
         year = 2003,
        month = oct,
       volume = {345},
       number = {1},
        pages = {340-348},
          doi = {10.1046/j.1365-8711.2003.06942.x},
archivePrefix = {arXiv},
       eprint = {1408.5890},
 primaryClass = {astro-ph.EP},
       adsurl = {https://ui.adsabs.harvard.edu/abs/2003MNRAS.345..340G}
}

@ARTICLE{georg2009,
       author = {{Georgakarakos}, Nikolaos},
        title = "{Improved equations for eccentricity generation in hierarchical triple systems}",
      journal = {\mnras},
         year = 2009,
        month = jan,
       volume = {392},
       number = {3},
        pages = {1253-1263},
          doi = {10.1111/j.1365-2966.2008.14143.x},
archivePrefix = {arXiv},
       eprint = {0811.1146},
 primaryClass = {astro-ph},
       adsurl = {https://ui.adsabs.harvard.edu/abs/2009MNRAS.392.1253G}
}

@ARTICLE{grant59,
   author = {{Grant}, G.},
    title = "{A Spectroscopic and Photometric Study of the Eclipsing System Lambda Tauri.}",
  journal = {\apj},
     year = 1959,
    month = jan,
   volume = 129,
    pages = {78},
      doi = {10.1086/146597},
   adsurl = {http://adsabs.harvard.edu/abs/1959ApJ...129...78G}
}

@ARTICLE{fotel1,
   author = {{Hadrava}, P.},
    title = "{Eclipsing binaries-light curve solutions}",
  journal = {Contributions of the Astronomical Observatory Skalnat\'e Pleso},
     year = 1990,
   volume = 20,
    pages = {23-25},
   adsurl = {http://adsabs.harvard.edu/cgi-bin/nph-bib_query?bibcode=1990CoSka..20...23H&db_key=AST}
}

@ARTICLE{fotel2,
   author = {{Hadrava}, P.},
    title = "{FOTEL 4 - User's guide}",
  journal = {Publ. Astron. Inst. Acad. Sci. Czech Rep.},
     year = 2004,
   volume = 92,
    pages = {1-14},
   adsurl = {http://adsabs.harvard.edu/cgi-bin/nph-bib_query?bibcode=2004PAICz..92....1H&db_key=AST}
}

@ARTICLE{gb1995,
       author = {{Gim{\'e}nez}, A. and {Bastero}, M.},
        title = "{A Revision of the Ephemeris-Curve Equations for Eclipsing Binaries with Apsidal Motion}",
      journal = {\apss},
         year = 1995,
        month = apr,
       volume = {226},
       number = {1},
        pages = {99-107},
          doi = {10.1007/BF00626903},
       adsurl = {https://ui.adsabs.harvard.edu/abs/1995Ap&SS.226...99G}
}

@ARTICLE{korel1,
   author = {{Hadrava}, P.},
    title = "{Orbital elements of multiple spectroscopic stars.}",
  journal = {\aaps},
     year = 1995,
    month = dec,
   volume = 114,
    pages = {393-+},
   adsurl = {http://cdsads.u-strasbg.fr/cgi-bin/nph-bib_query?bibcode=1995A%26AS..114..393H&db_key=AST}
}

@ARTICLE{korel2,
   author = {{Hadrava}, P.},
    title = "{Relative line photometry of eclipsing binaries.}",
  journal = {\aaps},
     year = 1997,
    month = may,
   volume = 122,
    pages = {581-584},
   adsurl = {http://cdsads.u-strasbg.fr/cgi-bin/nph-bib_query?bibcode=1997A%26AS..122..581H&db_key=AST}
}

@ARTICLE{korel3,
   author = {{Hadrava}, P.},
    title = "{KOREL - User's guide}",
  journal = {Publ. Astron. Inst. Acad. Sci. Czech Rep.},
     year = 2004,
    month = may,
   volume = 92,
    pages = {15-35},
   adsurl = {http://cdsads.u-strasbg.fr/cgi-bin/nph-bib_query?bibcode=1997A%26AS..122..581H&db_key=AST}
}

@ARTICLE{kostov2024,
       author = {{Kostov}, V.~B. and {Rappaport}, S.~A. and {Borkovits}, T. and {Powell}, B.~P. and {Gagliano}, R. and {Omohundro}, M. and {B{\'\i}r{\'o}}, I.~B. and {Moe}, M. and {Howell}, S.~B. and {Mitnyan}, T. and {Clark}, C.~A. and {Kristiansen}, M.~H. and {Terentev}, I.~A. and {Schwengeler}, H.~M. and {P{\'a}l}, A. and {Vanderburg}, A.},
        title = "{TIC 290061484: A Triply Eclipsing Triple System with the Shortest Known Outer Period of 24.5 Days}",
      journal = {\apj},
         year = 2024,
        month = oct,
       volume = {974},
       number = {1},
          eid = {25},
        pages = {25},
          doi = {10.3847/1538-4357/ad7368},
archivePrefix = {arXiv},
       eprint = {2410.01711},
 primaryClass = {astro-ph.SR},
       adsurl = {https://ui.adsabs.harvard.edu/abs/2024ApJ...974...25K}
}

@ARTICLE{mr,
       author = {{Harmanec}, Petr},
        title = "{Stellar Masses and Radii Based on Modern Binary Data}",
      journal = {Bulletin of the Astronomical Institutes of Czechoslovakia},
         year = 1988,
        month = dec,
       volume = {39},
        pages = {329},
       adsurl = {https://ui.adsabs.harvard.edu/abs/1988BAICz..39..329H}
}

@ARTICLE{hec2026,
       author = {{Harmanec}, P. and {Yang}, S. and {{\v{S}}lechta}, M. and {Grundstrom}, E.~D. and {Ribeiro}, J. and {Harmanec}, A.},
        title = "{The Be star o Cas is indeed the primary of a triple system}",
      journal = {\aap},
         year = 2026,
        month = may,
       volume = {709},
          eid = {A167},
        pages = {A167},
          doi = {10.1051/0004-6361/202659678},
archivePrefix = {arXiv},
       eprint = {2604.20747},
 primaryClass = {astro-ph.SR},
       adsurl = {https://ui.adsabs.harvard.edu/abs/2026A&A...709A.167H}
}

@ARTICLE{sef0,
   author = {{Horn}, J. and {Kub\'at}, J. and {Harmanec}, P. and
   {Koubsk\'y}, P. and {Hadrava}, P. and {\v{S}imon}, V. and {\v{S}tefl}, S.
   and {\v{S}koda}, P.},
    title = "{Spectroscopic orbit of the triple star 55 Ursae Majoris.}",
  journal = {\aap},
     year = 1996,
    month = may,
   volume = 309,
    pages = {521-529},
   adsurl = {http://adsabs.harvard.edu/cgi-bin/nph-bib_query?bibcode=1996A%26A...309..521H&db_key=AST}
}

@ARTICLE{john66,
   author = {{Johnson}, H.~L. and {Iriarte}, B. and {Mitchell}, R.~I. and
            {Wisniewski}, W.~Z.},
    title = "{UBVRIJKL photometry of the bright stars.}",
  journal = {Communications of the Lunar and Planetary Laboratory},
     year = 1966,
   volume = 4,
    pages = {99-110},
   adsurl = {http://adsabs.harvard.edu/abs/1966CoLPL...4...99J}
}

@INPROCEEDINGS{kaufer98,
       author = {{Kaufer}, Andreas and {Pasquini}, Luca},
        title = "{FEROS: the new fiber-linked echelle spectrograph for the ESO 1.52-m telescope}",
    booktitle = {Optical Astronomical Instrumentation},
         year = 1998,
       editor = {{D'Odorico}, Sandro},
       series = {Society of Photo-Optical Instrumentation Engineers (SPIE) Conference Series},
       volume = {3355},
        month = jul,
        pages = {844-854},
          doi = {10.1117/12.316798},
       adsurl = {https://ui.adsabs.harvard.edu/abs/1998SPIE.3355..844K}
}

@ARTICLE{spefo3,
   author = {{Krpata}, J.},
  journal = {http://astro.troja.mff.cuni.cz/ftp/hec/SPEFO/},
     year = 2008
}

@ARTICLE{mayor2003,
       author = {{Mayor}, M. and {Pepe}, F. and {Queloz}, D. and {Bouchy}, F. and {Rupprecht}, G. and {Lo Curto}, G. and {Avila}, G. and {Benz}, W. and {Bertaux}, J.-L. and {Bonfils}, X. and {Dall}, Th. and {Dekker}, H. and {Delabre}, B. and {Eckert}, W. and {Fleury}, M. and {Gilliotte}, A. and {Gojak}, D. and {Guzman}, J.~C. and {Kohler}, D. and {Lizon}, J.-L. and {Longinotti}, A. and {Lovis}, C. and {Megevand}, D. and {Pasquini}, L. and {Reyes}, J. and {Sivan}, J.-P. and {Sosnowska}, D. and {Soto}, R. and {Udry}, S. and {van Kesteren}, A. and {Weber}, L. and {Weilenmann}, U.},
        title = "{Setting New Standards with HARPS}",
      journal = {The Messenger},
         year = 2003,
        month = dec,
       volume = {114},
        pages = {20-24},
       adsurl = {https://ui.adsabs.harvard.edu/abs/2003Msngr.114...20M}
}

@ARTICLE{mcl37,
   author = {{McLaughlin}, D.~B.},
    title = "{Spectrographic studies opf eclipsing binaries}",
  journal = {Publications of Michigan Observatory},
     year = 1937,
   volume = 6,
    pages = {3},
   adsurl = {http://adsabs.harvard.edu/abs/1937POMic...6....1M}
}

@ARTICLE{nemr2013,
   author = {{Nemravov{\'a}}, J.~A. and {Harmanec}, P. and {Bencheikh}, J. and
{Bolton}, C.~T. and {Bo{\v z}i{\'c}c}, H. and {Bro{\v z}}, M. and
{Engle}, S. and {Grunhut}, J. and {Guinan}, E.~F. and {Hummel}, C.~A. and
{Kor{\v c}{\'a}kov{\'a}}, D. and {Koubsk{\'y}}, P. and {Mayer}, P. and
{Mourard}, D. and {Ribeiro}, J. and {{\v S}lechta}, M. and {Vokrouhlick{\'y}}, D. and
{Votruba}, V. and {Wolf}, M. and {Zasche}, P. and {Chara/Vega Team} and
{Npoi Team}},
    title = "{An Unusual Quadruple System {$\xi$} Tauri}",
  journal = {Central European Astrophysical Bulletin},
     year = 2013,
   volume = 37,
    pages = {207-216},
   adsurl = {http://adsabs.harvard.edu/abs/2013CEAB...37..207N}
}

@ARTICLE{nemr2016,
       author = {{Nemravov{\'a}}, J.~A. and {Harmanec}, P. and {Bro{\v{z}}}, M. and {Vokrouhlick{\'y}}, D. and {Mourard}, D. and {Hummel}, C.~A. and {Cameron}, C. and {Matthews}, J.~M. and {Bolton}, C.~T. and {Bo{\v{z}}i{\'c}}, H. and {Chini}, R. and {Dembsky}, T. and {Engle}, S. and {Farrington}, C. and {Grunhut}, J.~H. and {Guenther}, D.~B. and {Guinan}, E.~F. and {Kor{\v{c}}{\'a}kov{\'a}}, D. and {Koubsk{\'y}}, P. and {K{\v{r}}{\'\i}{\v{c}}ek}, R. and {Kuschnig}, R. and {Mayer}, P. and {McCook}, G.~P. and {Moffat}, A.~F.~J. and {Nardetto}, N. and {Pr{\v{s}}a}, A. and {Ribeiro}, J. and {Rowe}, J. and {Rucinski}, S. and {{\v{S}}koda}, P. and {{\v{S}}lechta}, M. and {Tallon-Bosc}, I. and {Votruba}, V. and {Weiss}, W.~W. and {Wolf}, M. and {Zasche}, P. and {Zavala}, R.~T.},
        title = "{{\ensuremath{\xi}}Tauri: a unique laboratory to study the dynamic interaction in a compact hierarchical quadruple system}",
      journal = {\aap},
         year = 2016,
        month = oct,
       volume = {594},
          eid = {A55},
        pages = {A55},
          doi = {10.1051/0004-6361/201628860},
       adsurl = {https://ui.adsabs.harvard.edu/abs/2016A&A...594A..55N}
}

@ARTICLE{nik38,
   author = {{Nikonov}, V.~B.},
    title = "{Photoelectric observations of the eclipsing binary lambda Tau}",
  journal = {Bull. Abastumani Astrophys. Obs.},
     year = 1938,
   volume = 2,
    pages = {29-36},
}

@BOOK{esa97,
    author = {{Perryman}, M.~A.~C. and {ESA}},
    title = "{The HIPPARCOS and TYCHO catalogues}",
publisher = {The Hipparcos and Tycho catalogues.~Astrometric and
photometric star catalogues derived from the ESA Hipparcos Space
Astrometry Mission, Publisher: Noordwijk, Netherlands:
ESA Publications Division, 1997, Series:
ESA SP Series 1200},
year = 1997,
   adsurl = {http://cdsads.u-strasbg.fr/cgi-bin/nph-bib_query?bibcode=1997hity.book.....P&db_key=AST}
}

@ARTICLE{pribulla2024,
       author = {{Pribulla}, T. and {Va{\v{n}}ko}, M. and {Kom{\v{z}}{\'\i}k}, R. and {Sivani{\v{c}}}, P.},
        title = "{High-resolution {\'e}chelle spectrograph at Skalnat{\'e} Pleso Observatory}",
      journal = {Contributions of the Astronomical Observatory Skalnate Pleso},
         year = 2024,
        month = feb,
       volume = {54},
       number = {2},
        pages = {43-46},
          doi = {10.31577/caosp.2024.54.2.43},
       adsurl = {https://ui.adsabs.harvard.edu/abs/2024CoSka..54b..43P}
}

@ARTICLE{prsa2005,
   author = {{Pr{\v s}a}, A. and {Zwitter}, T.},
    title = "{A Computational Guide to Physics of Eclipsing Binaries.
              I. Demonstrations and Perspectives}",
  journal = {\apj},
   eprint = {astro-ph/0503361},
     year = 2005,
    month = jul,
   volume = 628,
    pages = {426-438},
      doi = {10.1086/430591},
   adsurl = {http://adsabs.harvard.edu/cgi-bin/nph-bib_query?bibcode=2005ApJ...628..426P&db_key=AST}
}

@ARTICLE{prsa2006,
   author = {{Pr{\v s}a}, A. and {Zwitter}, T.},
    title = "{Disentangling Effective Temperatures of Individual Eclipsing
              Binary Components by Means of Color-Index Constraining}",
  journal = {\apss},
   eprint = {astro-ph/0602205},
     year = 2006,
    month = jul,
    pages = {36-+},
      doi = {10.1007/s10509-006-9154-4},
   adsurl = {http://adsabs.harvard.edu/cgi-bin/nph-bib_query?bibcode=2006Ap%26SS.tmp...36P&db_key=AST}
}

@ARTICLE{rappaport2026,
       author = {{Rappaport}, S.~A. and {Borkovits}, T. and {Torres}, G. and {Chen}, H.-L. 
	       and {Zasche}, P. and {Vokrouhlick{\'y}}, D. and {Deme}, B. and 
	       {Ma{\v{s}}ek}, M. and {Ku{\v{c}}{\'a}kov{\'a}}, H. and {Uhla{\v{r}}}, R. and 
	       {Wolf}, M. and {Garai}, Z. and {Pribulla}, T. and {B{\'\i}r{\'o}}, I.~B. and 
	       {Cs{\'a}nyi}, I. and {Dencs}, Z. and {Latham}, D. and {Mitnyan}, T. and 
	       {Kostov}, V.~B. and {Powell}, B.~P.},
        title = "{TIC 366240660 - A dynamically active triple with a very short outer period of 30.5 days}",
      journal = {\aap},
         year = 2026,
        month = oct,
       volume = {714},
          eid = {A523},
        pages = {A523},
          doi = {10.1051/0004-6361/202662373},
       adsurl = {https://ui.adsabs.harvard.edu/abs/2026A&A...714A.523R}
}

@ARTICLE{rens79,
   author = {{Renson}, P. and {Manfroid}, J.},
    title = "{New Research for Periods of Ap Stars Observed at the ESO - III}",
  journal = {Information Bulletin on Variable Stars},
     year = 1979,
    month = aug,
   volume = 1658,
    pages = {1},
   adsurl = {http://cdsads.u-strasbg.fr/abs/1979IBVS.1658....1R}
}

@ARTICLE{schle16a,
   author = {{Schlesinger}, F.},
    title = "{The orbit of [lambda] Tauri}",
  journal = {Publ. Allegheny Obs.},
     year = 1916,
   volume = 3,
    pages = {23-30},
   adsurl = {http://adsabs.harvard.edu/abs/1916PAllO...3...23S}
}

@ARTICLE{schle16b,
   author = {{Schlesinger}, F.},
    title = "{Spectrographic observations of [lambda] Tauri, indicating the presence of a third body in the system}",
  journal = {Publ. Allegheny Obs.},
     year = 1916,
   volume = 3,
    pages = {167-178},
   adsurl = {http://adsabs.harvard.edu/abs/1916PAllO...3..167S}
}

@ARTICLE{schmidt1858,
       author = {{Schmidt}, Johann Friedrich Julius},
        title = "{{\"U}ber ver{\"a}nderliche Sterne, von Herrn J. F. Julius Schmidt}",
      journal = {Astronomische Nachrichten},
         year = 1858,
        month = feb,
       volume = {47},
       number = {24},
        pages = {373},
          doi = {10.1002/asna.18580472406},
       adsurl = {https://ui.adsabs.harvard.edu/abs/1858AN.....47..373S}
}

@ARTICLE{sepin2007,
       author = {{Sepinsky}, J.~F. and {Willems}, B. and {Kalogera}, V. and {Rasio}, F.~A.},
        title = "{Interacting Binaries with Eccentric Orbits: Secular Orbital Evolution Due to Conservative Mass Transfer}",
      journal = {\apj},
         year = 2007,
        month = oct,
       volume = {667},
       number = {2},
        pages = {1170-1184},
          doi = {10.1086/520911},
archivePrefix = {arXiv},
       eprint = {0706.4312},
 primaryClass = {astro-ph},
       adsurl = {https://ui.adsabs.harvard.edu/abs/2007ApJ...667.1170S}
}

@INPROCEEDINGS{spefo,
   author = {{\v{S}koda}, P.},
    title = "{SPEFO---A Simple, Yet Powerful Program for
              One-Dimensional Spectra Processing}",
booktitle = {ASP Conf. Ser. 101: Astronomical Data Analysis Software and Systems V},
     year = 1996,
    pages = {187-189},
   adsurl = {http://adsabs.harvard.edu/cgi-bin/nph-bib_query?bibcode=1996ASPC..101..187S&db_key=AST}
}

@INPROCEEDINGS{vokorel,
       author = {{{\v{S}}koda}, Petr and {Hadrava}, Petr and {Fuchs}, Jan},
        title = "{VO-KOREL: A Fourier Disentangling Service of the Virtual Observatory}",
    booktitle = {From Interacting Binaries to Exoplanets: Essential Modeling Tools},
         year = 2012,
       editor = {{Richards}, Mercedes T. and {Hubeny}, Ivan},
       series = {IAU Symposium},
       volume = {282},
        month = apr,
        pages = {403-404},
          doi = {10.1017/S1743921311027931},
archivePrefix = {arXiv},
       eprint = {1112.2784},
 primaryClass = {astro-ph.IM},
       adsurl = {https://ui.adsabs.harvard.edu/abs/2012IAUS..282..403S}
}

@ARTICLE{sordel75,
       author = {{Soderhjelm}, S.},
        title = "{The three-body problem and eclipsing binaries. Application to Algol and lambda Tauri.}",
      journal = {\aap},
         year = 1975,
        month = aug,
       volume = {42},
        pages = {229-236},
       adsurl = {https://ui.adsabs.harvard.edu/abs/1975A&A....42..229S}
}

@ARTICLE{steb20,
   author = {{Stebbins}, J.},
    title = "{The eclipsing variable star lambda Tauri.}",
  journal = {\apj},
     year = 1920,
    month = may,
   volume = 51,
    pages = {193-217},
      doi = {10.1086/142540},
   adsurl = {http://adsabs.harvard.edu/abs/1920ApJ....51..193S}
}

@ARTICLE{steb20b,
   author = {{Stebbins}, J.},
    title = "{The ellipsoidal variable star pi.05 Orionis.}",
  journal = {\apj},
     year = 1920,
    month = may,
   volume = 51,
    pages = {218-222},
      doi = {10.1086/142541},
   adsurl = {http://adsabs.harvard.edu/abs/1920ApJ....51..218S}
}

@ARTICLE{talens2018,
       author = {{Talens}, G.~J.~J. and {Deul}, E.~R. and {Stuik}, R. and {Burggraaff}, O. and {Lesage}, A.-L. and
	       {Spronck}, J.~F.~P. and {Mellon}, S.~N. and {Bailey}, J.~I. and {Mamajek}, E.~E. and
	       {Kenworthy}, M.~A. and {Snellen}, I.~A.~G.},
        title = "{Data calibration for the MASCARA and bRing instruments}",
      journal = {\aap},
         year = 2018,
        month = nov,
       volume = {619},
          eid = {A154},
        pages = {A154},
          doi = {10.1051/0004-6361/201834070},
archivePrefix = {arXiv},
       eprint = {1810.04060},
 primaryClass = {astro-ph.EP},
       adsurl = {https://ui.adsabs.harvard.edu/abs/2018A&A...619A.154T}
}

@INPROCEEDINGS{tody93,
       author = {{Tody}, Doug},
        title = "{IRAF in the Nineties}",
    booktitle = {Astronomical Data Analysis Software and Systems II},
         year = 1993,
       editor = {{Hanisch}, R.~J. and {Brissenden}, R.~J.~V. and {Barnes}, J.},
       series = {Astronomical Society of the Pacific Conference Series},
       volume = {52},
        month = jan,
        pages = {173},
       adsurl = {https://ui.adsabs.harvard.edu/abs/1993ASPC...52..173T}
}

@ARTICLE{toko2013,
       author = {{Tokovinin}, Andrei and {Fischer}, Debra A. and {Bonati}, Marco and {Giguere}, Matthew J. and {Moore}, Peter
	       and {Schwab}, Christian and {Spronck}, Julien F.~P. and {Szymkowiak}, Andrew},
        title = "{CHIRON{\textemdash}A Fiber Fed Spectrometer for Precise Radial Velocities}",
      journal = {\pasp},
         year = 2013,
        month = nov,
       volume = {125},
       number = {933},
        pages = {1336},
          doi = {10.1086/674012},
archivePrefix = {arXiv},
       eprint = {1309.3971},
 primaryClass = {astro-ph.IM},
       adsurl = {https://ui.adsabs.harvard.edu/abs/2013PASP..125.1336T}
}

@ARTICLE{aret13,
   author = {{von Aretin}, E.~F.},
    title = "{Untersuchungen uber den Lichtwechsel und die spektroskopische Bahn von [ lambda] Tauri}",
  journal = {Astronomische Mitteilungen der Universitaets-Sternwarte zu Goettingen},
     year = 1913,
   volume = 15,
    pages = {1},
   adsurl = {http://adsabs.harvard.edu/abs/1913MiGoe..15....1V}
}

@ARTICLE{wra2012,
   author = {{Wraight}, K.~T. and {Fossati}, L. and {Netopil}, M. and {Paunzen}, E. and
{Rode-Paunzen}, M. and {Bewsher}, D. and {Norton}, A.~J. and
{White}, G.~J.},
    title = "{A photometric study of chemically peculiar stars with the STEREO satellites - I. Magnetic chemically peculiar stars}",
  journal = {\mnras},
archivePrefix = "arXiv",
   eprint = {1110.6283},
 primaryClass = "astro-ph.SR",
     year = 2012,
    month = feb,
   volume = 420,
    pages = {757-772},
      doi = {10.1111/j.1365-2966.2011.20090.x},
   adsurl = {http://cdsads.u-strasbg.fr/abs/2012MNRAS.420..757W}
}

@ARTICLE{yin2025,
       author = {{Yin}, Jia and {Yao}, Yongqiang and {Qian}, Xuan and {Liu}, Liyong and {Chen}, Xu and {Zhai}, Liuming},
        title = "{Calibration and applications of the all-sky camera at the Ali Observatory in Tibet}",
      journal = {\mnras},
         year = 2025,
        month = feb,
       volume = {537},
       number = {1},
        pages = {617-627},
          doi = {10.1093/mnras/staf056},
archivePrefix = {arXiv},
       eprint = {2501.08358},
 primaryClass = {astro-ph.IM},
       adsurl = {https://ui.adsabs.harvard.edu/abs/2025MNRAS.537..617Y}
}

@ARTICLE{zen83,
   author = {{Zentelis}, N.},
    title = "{Radial velocities for early type stars in six galactic regions}",
  journal = {\aaps},
     year = 1983,
    month = sep,
   volume = 53,
    pages = {445-453},
   adsurl = {http://adsabs.harvard.edu/abs/1983A%26AS...53..445Z}
}

\begin{appendix}
\nolinenumbers
\section{Overview of available spectroscopic observations}\label{apa}

The initial reductions of all electronic spectra were performed using the IRAF
and SPEFO programmes.\footnote{\texttt{IRAF} is distributed by the~National
Optical Astronomy Observatories, operated by the~Association of Universities
for Research in Astronomy, Inc., under contract to the~National Science
Foundation of the~United States \citep{tody93};
SPEFO was originally written by the late J.~Horn and later further
developed by J.~Krpata \citep{sef0,spefo,spefo3}.}
Spectra normalization, and cleaning from cosmics and flaws
were carried out with the Java program \respefoe2
\citep[see the appendix in][]{hec2026}. Interactive RV measurements, based
on the comparison of direct and flipped line profiles, were also carried
out in \respefoe2, independently by Petr Harmanec and Pavel Dole\v{z}al
to obtain some estimates of measuring errors.

We carried out a careful compilation of all available RV measurements of \lat
with known dates of observations from the literature.
These were complemented by our own RV measurements.
In cases when the original sources give the dates and times of mid-exposures
we used the programme HEC19 to derive heliocentric Julian
dates (HJDs).\footnote{This programme, which
handles data of various forms, together with brief instructions on how
to use it, is freely available on the anonymous ftp server at
\url{http://astro.troja.mff.cuni.cz/ftp/hec/HEC19}}
The log of all RVs used in this study is in Table~\ref{jourv}, while
the log of all electronic spectra at our disposal is in Table~\ref{jousp}.
For convenience of future investigators, we publish all mean individual RVs, both
from the literature and measured by us in (electronic only) Table~A.1.

Some comments on individual data files follow. The files are identified
by the spectrograph Nos. given in Tables~\ref{jourv} and \ref{jousp}.

\begin{itemize}
\item Spg. 1: \citet{belo1898} tabulated Pulkowa times and RVs
in geographic miles. We converted his mid-exposure times to
HJDs noting that G.M.T. = Pulkovo time - 2.0219 h.
Using a Belopolsky's paper on $\beta$~Lyr, where he tabulated RVs in both,
the geographical miles and \ks, we found that he adopted a conversion
factor 1~geographic mile =  7.4211 km and used it to convert his RVs
to \ks. He reported two RVs for the secondary, but these are quite out of
the secondary RV curve, and we omit them.
\item Spg. 2: \citet{aret13} published a very detailed analysis
of 23 prismatic spectra from the Potsdam refractor. Since he tabulated
his RV measurements for the comparison and stellar lines (carried out
twice as measurements A and B, respectively, reversing the position of
the plates on the comparator), we were able to recalculate the wavelength fits
and derive RVs of stellar lines using modern wavelengths.
When doing so, we noted the following obvious misprints in the
tabulation: On page 106 in `measurement~B' the comparison lines
14 and 15 are actually 15 and 16. On page 112, measurement A, the stellar line
13 is actually line 11. On page 114, measurement A, the position of line 7
should be 22.667 (not 27.667 as printed). In principle, up to 14 stellar
lines of \ion{H}{i}, \ion{He}{i}, and \ion{Mg}{ii} between 3889 and 4923~\AA\
could be measured. However, depending on the plate quality, only some of them
were actually measured on each particular plate. With two repeated
measurements, this constitute up to 28 individual RVs for each plate.
Therefore, we derived robust mean values for each plate and using the
corresponding root-mean-square (rms) errors, we derived plate weight
inversely proportional to (rms)$^2$.
\item Spgs. 3\&4: \citet{schle16a,schle16b} published mid-exposures and
RVs for 126 spectra obtained at the Allegheny Observatory. We used his
mid-exposure times to derive the corresponding HJDs.
Of these spectrograms, 98 were later remeasured by \citet{ebbig56}, who used
modern laboratory wavelengths. They also measured four earlier Allegheny
spectra, not measured by Schlesinger, and 5 other, obtained after
Schlesinger published his studies. The total number of spectra measured by
\citet{ebbig56} is thus 107. There are 28 spectra, which were not
re-measured by \citet{ebbig56} and for which we have to use the
original Schlesinger RVs, based on older wavelengths. That is, why we
assigned them a separate spg. number 3 to allow for a small difference
in the velocity zero point with respect to spg. 4 with RVs based on
\citet{ebbig56} RV measurements. We note, however, that there is no
large difference between the original and re-measured RV values.
\item Spg. 5: \citet{mcl37} published RVs measured on the 1-prism spectra
from Ann Arbor Observatory. We just derived HJDs from the published dates
of observations and adopted his measured values.
\item Spg. 6: RVs measured on the prismatic spectra from the 1.83~m reflector
of the Dominion Astrophysical Observatory were adopted from the study
by \citet{ebbig56}.
\item Spg. 7, 8, and 9: \citet{ebbig56} obtained 2~Lick spectra
during two observing nights but published only mean RVs for each night. Similarly,
they also published 68 nightly normal RVs based on RVs from 350 Mt.~Wilson
spectra. There is no way to distinguish Mt.~Wilson and Lick spectra
in their Table~5 so we assign the same zero point to them. In addition, we also
append a~single RV published by \citet{zen83}.
\item Spg. 10: \citet{fekel82} used the McDonald 2.7-m coud\'e
spectrograph and a Reticon detector with 1024 pixels. The covered
spectral region was 105~\AA\ centred on 4510~\AA. The resolution was
0.45~\AA\ = 30 \kms per 1 pixel, and most of the spectra had a
$S/N$ ratio of $400-500$. \citet{fekel82} published individual
\ion{He}{I}~4471~\AA\ and \ion{Mg}{II}~4481~\AA\ RVs for the primary,
and \ion{Mg}{II}~4481~\AA\ RVs and mean RVs of several \ion{Fe}{II}
and \ion{Ti}{II} lines for the secondary.
\item Spg. 11: We have at our disposal only the red segments of the
CCD spectra from the CORALIE echelle spectrograph attached to the ESO LaSilla
Euler Swiss 1.2~m telescope.
\item Spg. 12: CCD spectra from the FEROS echelle spectrograph attached
to the ESO LaSilla 2.2~m Cassegrain reflector \citep{kaufer98} were downloaded
and preliminary reduced, rectified, and cleaned from cosmics by Jana \v{S}vr\v{c}kov\'a.
\item Spg. 13: HARPS CCD spectra from the echelle spectrograph attached
to the ESO LaSilla 3.6~m reflector \citep{mayor2003} were also downloaded
and preliminary reduced by Jana \v{S}vr\v{c}kov\'a.
\item Spg. 14: Red and a few blue spectra obtained with a coud\'e CCD spectrograph
attached to the 2.0~m Ond\v{r}ejov reflector were secured by various observers. They
were calibrated in wavelength and converted to 1-D frames by Miroslav \v{S}lechta in IRAF.
\item Spg. 15: Red CCD spectra obtained with the Littrow LHIRES~III spectrograph
attached to the Lisbon 0.356~m reflector were secured and preliminary reduced by
Jose Ribeiro.
\item Spg. 16: Two dense series of CHIRON echelle
spectra \citep{toko2013}, both covering the whole 33~d outer period,
were secured by us at the CTIO in 2021 and 2022.
Their preliminary reduction (wavelength calibration and
creation of 1-D images) was carried out with the pipeline
of the CTIO Observatory. Their further reduction, i.e. the order-by-order
normalisation and creation of combined complete spectra with properly treated
overlap between the blue orders, was carried out with the programme
{\tt resimplex} written by Pavel Dole\v{z}al
\footnote{\url{https://github.com/pavel-dolezal/resimplex}}
\citep[see also Appendix~A in][]{broz2026}.
The removal of cosmic spikes and occasional flaws
and preliminary RV measurements were carried out with the programme \respefoe2
\footnote{\url{https://astro.troja.mff.cuni.cz/projects/respefo/}}.
\item Spg. 17: Skalnat\'e Pleso 1.3 m Nasmyth-Cassegrain reflector with an echelle
spectrograph \citep{pribulla2024} was used. The preliminary reduction of the spectra
was carried out by Theo Pribulla.
\end{itemize}

\section{Photometry}\label{apb}

Photoelectric observations used in this study consist of observations
secured in non-standard systems, \ubv\ observations, and $BVRI$ observations.
Their overview is in Table~\ref{jouphot}, and in the following we provide
details on individual datasets.
There are also test observations from the MASCARA survey \citep{talens2018}
and from a CCD camera from the Ali Observatory in Tibet \citep{yin2025}. These later
were kindly put at our disposal by Jia Yin. They show eclipses in proper phases,
but their scatter is rather high, and we do not use them here.
All individual observations are provided in electronic only Tables B.1
to B.8 deposited to CDS.

\begin{itemize}
\item File A: \citet{bax1848} \ \ Accurate times of these discovery
visual observations are not known. We only know that Baxendell observed
brightness decreases of \lat on the nights 1848~Nov.~6, 10, 14, 18, and 22.
He mentions that the brightness of \lat was comparable to that
of $\nu$~Tau = HD~25490 while it was much brighter than it on
1848~Nov~7. Since $\nu$~Tau has $V=3$\m9, it is clear that Baxendell indeed
observed the primary minima on the five nights mentioned above.
\item File B and Table B.1: \citet{steb20} \ \
These are the first photoelectric observations
of \lat obtained with a~photoelectric cell designed by Jakob Kunz and attached
to 0.305~m refractor of University of Illinois Observatory.
Stebbins observed \lat alternatively relative to five different comparison stars:
$\xi$~Tau = HD~21364 (which is now known to be a 7\fd1467 eclipsing binary),
$\pi^5$~Ori = HD~31237 (which Stebbins himself identified as an ellipsoidal
variable with a 3\fd70045 period),
$\pi^4$~Ori = HD~30836, e~Tau = 30~Tau = HD~23793 \citep[which is
suspected of variability under name NSV~1327 but no obvious variations are
seen in the Hipparcos $H_{\rm p}$ photometry by][]{esa97},
and $\mu$~Tau = HD~26912. For observations relative to 30~Tau and $\mu$~Tau,
Stebbins attenuated the light of \lat by attenuating filters. He, however,
measured the amount of this attenuation as well as the mean magnitude
differences between individual comparison stars. Moreover, he corrected his
observations relative to $\pi^5$~Ori for its known periodic light variations.
After all these corrections, he tabulated all magnitude differences relative
to $\xi$~Tau versus HJDs.
As a precaution, we treated observations of \lat relative
to those five comparisons as separate datasets and allowed \phoebe to derive
individual zero points for them.
We omitted the following observations: Observations relative to $\xi$~Tau:
HJD~2421252.522 and .542  (near primary minima of $\xi$~Tau),
HJD~2421257.577, 2421269.547, and 2421534.799  (near secondary minima
of $\xi$~Tau); one observation relative to e~Tau:  HJD 2421529.778, and
one observations relative to $\pi^4$~Ori: HJD~2421296.541 (poor sky).

We note that the numbers of observations in Table~\ref{jouphot} give the
original number of observations as well as the actual number of observations,
which we used.
\item File C and Table B.2: \citet{nik38} \ \  These first photoelectric observations
secured with photomultipliers were obtained with a Guthnick photometer
at Abastumani in seasons 1935 and 1936, and on MtCanobili with a Russian
photometer in 1937. No colour filters were used and the photomultipliers
clearly have the maximum of sensitivity in the blue region.
$\mu$~Tau (HD~26912) and $\nu$~Tau (HD~25490) served alternatively
as comparisons. All 1935 observations were obtained relative to $\mu$~Tau.
Because of (an unconfirmed) suspicion that $\mu$~Tau could be a~$\beta$~CMa
variable, Nikonov finally reduced all observations of \lat relative to
$\nu$~Tau. Since the mean measured magnitude difference $\mu$~Tau - $\nu$~Tau
were $-0$\m248$\pm0$\m002 in 1935 but $-0$\m277$\pm0$\m003 in 1936 (no value
was given for 1937), we treat the observations from the three seasons as
separate datasets. We omitted the following deviating observations:
2428060.444, all three from 28069, 28100.466, all from the
nights 28113, 28114, and 28115 (clouds), and 28437.545. There are also
two obvious typographical errors; observation 2428100.466 on page 32 should
have a magnitude difference 0.271, not 0.371, and on page 34 of the paper
the correct HJD of the first observation should read as 2428815.506\,.
\item File D and Table B.3: \citet{grant59} \ \  These observations were
obtained with a~photoelectric photometer attached to the
McDonald 0.33~m reflector. These are the first systematic
observations covering phases of both minima, which were secured in \ubv\
and transformed to the standard system by the author. Unfortunately, they
are tabulated relative to HR~1194 = HD~24155, which is now known as
a CP2 variable V766~Tau. There is an obvious typographical error on page 84
of the paper: The correct HJD should be 2435101.656 (not .756 as printed).
Besides, all $U$ observations for the nights HJD 2435132 and 2435144 are
printed with incorrect sign. However, the greatest puzzle is that if one
adds the observed $V$ magnitude of HR~1194 to the magnitude differences
\lat $-$ HR~1194, \lat appears as a sixth magnitude star! Conceivably,
Grant used some attentuating filter when observing \lat but no such
piece of information can be found in the paper. As it follows from a note
on the first page of his paper, Grant died before the paper was finished
and somebody else was preparing it for publication. There is no clue even
in his PhD Thesis. At the end, we simply assumed that Grant used a 3\m0
attenuating filter. Then his observations agree well with other datasets.
\item File E and Table B.4: \citet{esa97} \ \  Hipparcos \hp\ photometry
with error flags 0 and 1 was used.
\item File F:  STEREO photometry \citep{bewsher2010} represents a~very
rich but relatively noisy dataset. For that reason we only used it to confirm
our linear ephemeris derived from the global solution for all other photometric
data.
\item File G and Table B.5: this paper \ \ Photoelectric \ubvr\ photometry of the star
have been collected with a photometer attached to the 0.65~m reflector of
Hvar Observatory in Croatia. The data were carefully reduced to the standard
system via frequent observations of Johnson standards and non-linear
transformations and are tabulated for HJDs.
\item File H and Table B.6: this paper \ \ A series of $BVRI$ CCD observations of \lat
relative to HR~1243 = HD~25330 was obtained by Robert Uhl\'a\v{r} at
his private observatory at J\'\i lov\'e u~Prahy with a G2/KAF0402ME CCD
photometer attached to a 0.075 m reflector. These observations were reduced
and transformed to the standard Johnson $BV$ and Cousins $RI$ filters using
the C-MUNIPACK software.\footnote{\url{http://c-munipack.sourceforge.net}}
\item File I and Table B.7: MOST 2016 and 2017. These observations were downloaded
at the Vienna tracking station, reduced by Thomas Kallinger and tabulated versus HJDs.    
\item File J and Table B.8: this paper \ \  TESS photometry from sectors 05, 32,
42, 43, 44, 70, and 71 was downloaded and converted to magnitudes, using
an~arbitrary zero point, close to Johnson $V$ magnitude. Barycentric BJDs are tabulated for
these observations.
\item File K: This all-sky RGB photometry was obtained as a test case
during commissioning the Ali Observatory in Tibet by \citet{yin2025} and was kindly
put at our disposal by the author. We did not use it directly in our global
photometric solution but only to confirmation of our new linear ephemeris.
\end{itemize}

\begin{figure}
\resizebox{\hsize}{!}{\includegraphics{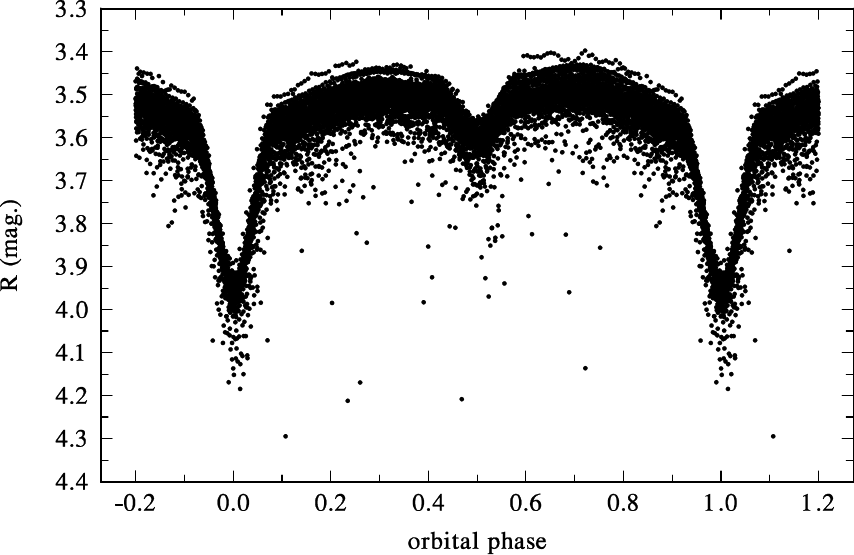}}
\resizebox{\hsize}{!}{\includegraphics{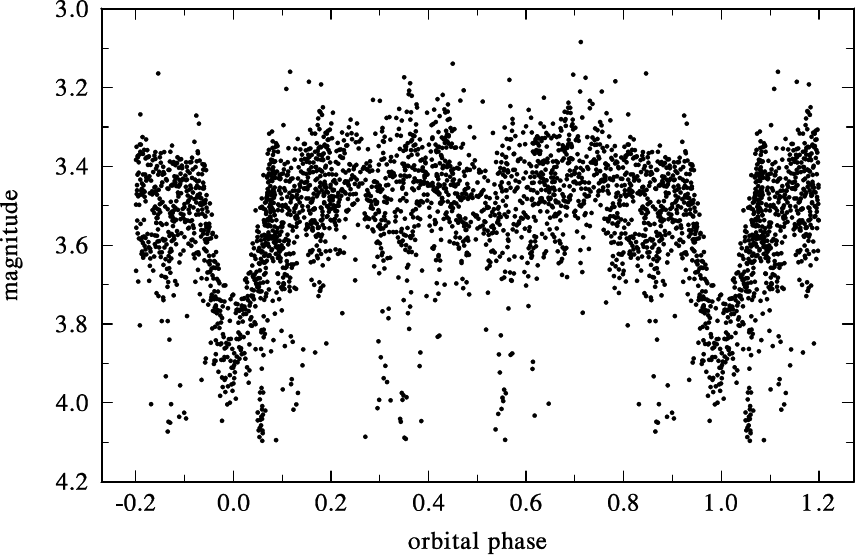}}
\caption{Phase plots using ephemeris~(\ref{efe2}) for data excluded
from our analyses. \textit{Top}: File F STEREO photometry.
\textit{Bottom}: File K photometry from Tibet.}\label{excluded}
\end{figure}

\end{appendix}

\end{document}